\documentstyle[12pt]{article}

\date{\today}

\def\be{\begin{equation}}
\def\ee{\end{equation}}
\def\bear{\begin{eqnarray}}
\def\eear{\end{eqnarray}}

\def\half{{{1\over 2}}}

\def\wdg{{\wedge}}                             

\def\Re{{\rm Re\hskip0.1em}}

\def\a{{\alpha}}
\def\b{{\beta}}
\def\g{{\gamma}}

\def\r{{\rho}}

\def\z{{\zeta}}

\def\lam{{\lambda}}

\def\e{{\epsilon}}

\def\er{{\epsilon_R}}

\def\ce{{ {\cal E}^{eff} }}
\def\cer{{ {\cal E}_R^{eff} }}

\def\vth{{\vartheta}}

\def\ba{{\overline{a}}}
\def\by{{\overline{y}}}

\def\bn{{\overline{n}}}
\def\bp{{\overline{p}}}
\def\bj{{\overline{j}}}

\def\bW{{\overline{W}}}
\def\bS{{\overline{S}}}

\def\bT{{\overline{T}}}

\def\td{{\tilde{d}}}
\def\cA{{\cal{A}} }
\def\hcA{{\hat {\cal{A}} }}
\def\hI{{\hat{I}} }
\def\hK{{\hat{K}} }
\def\hJ{{\hat{J}} }
\def\cR{{\cal{R}} }
\def\cO{{\cal{O}} }

\def\cS{{\cal{S}} }
\def\cX{{\cal{X}} }

\def\bm{{\overline{m}}}
\def\bn{{\overline{n}}}
\def\bp{{\overline{p}}}

\def\bC{{\overline{C}}}
\def\bZ{{\overline{Z}}}
\def\k{{\kappa_{11}^2}}
\begin{document}

%

\begin{titlepage}
\titlepage
\rightline{hep-th/0012104}
\rightline{ RUPT-????}
\rightline{\today}
\vskip 1cm

\centerline{{\Huge Instabilities in heterotic M-theory}}
\centerline{{\Huge induced by open membrane instantons.}}
\vskip 1cm
\centerline{Gregory Moore$^{\dagger}$, 
Grigor Peradze$^{\dagger  \diamond}$, and
Natalia Saulina
$^{\dagger \dagger}$}
\vskip 0.5cm
\begin{tabular}{cc}
$\dagger$ Department of Physics,
 & ~~~~~~~~~~~~$\diamond$ Department of Physics, \\
 Rutgers University &~~~~~~~~~~~~  Yale University \\
Piscataway, NJ 08854, USA &~~~~~~~~~~~~New Haven,CT 06520,USA\\
gmoore@physics.rutgers.edu &~~~~~~~~~~~~peradze@physics.rutgers.edu\\
\end{tabular}

\begin{center}
$\dagger \dagger$ Department of Physics, Jadwin Hall \\
Princeton University \\
NJ 08544, USA\\
saulina@princeton.edu
\end{center}

\abstract{We study the effective low energy supergravity of the 
strongly coupled heterotic string compactified on a Calabi-Yau 
3-fold with generic $E_8 \times E_8$ gauge bundle. 
We focus on the effective potential for the chiral scalars. 
The effective superpotential 
can be studied using the dual 11-dimensional M-theory background
involving  insertions of M5 branes along an interval. 
In such backgrounds, in some  regions of moduli space, the 
leading nonperturbative contributions are due to open membrane 
instantons.  These instantons   lead to both attractive and repulsive 
forces between the 5-branes and the orientifold ``M9-branes,'' 
depending on the region of moduli space. The resulting 
dynamics on moduli space include a strong coupling 
dual to the Dine-Seiberg instability, in which the interval 
grows. We discuss conditions under which the 5-branes 
are attracted to the wall and comment on the relevance of these results to 
the study of chirality-changing phase transitions in heterotic 
M-theory.}

\end{titlepage}
\tableofcontents

\section{Introduction}

In the past few years there have been significant advances in 
the study of strongly coupled heterotic string theory, thanks 
to the formulation in terms of $M$-theory on an interval 
$S^1/Z_2$ \cite{HorWit,HorWit1}. 
In particular, the compactification of $M$-theory on 
a product  of an interval with a Calabi-Yau
 3-fold (denoted hereafter by $\cX$ )
leads to qualitatively different 
physics from that of the weakly coupled heterotic 
string, as first noted in \cite{Witten,Dine, Hor}. 

In heterotic string compactification one must choose 
an instanton configuration for gauge fields along 
$\cX$. The so-called ``standard embedding'' identifies 
the gauge field with the spin connection of the metric. 
Other choices of gauge instantons, the so-called 
 ``nonstandard 
embeddings,'' are closely related, in the strongly coupled 
regime, to backgrounds obtained by including insertions of M5-branes
wrapping 
a product of 4-dimensional spacetime with a
holomorphic curve $\Sigma$  in $\cX$. 
At low energies, the 
physics of such backgrounds is summarized by a 
complicated $d=4,N=1$ supergravity theory. 
It has been shown in 
\cite{Ovrut1}-\cite{Ovrut5} 
that such  backgrounds can lead to 
phenomenologically interesting gauge groups, and 
it is therefore of interest to understand more 
completely the full low energy supergravity in such 
backgrounds. 
While several 
aspects of the effective Lagrangian have been 
worked out in \cite{Witten,Dine},\cite{Ovrut1}-\cite{Ovrut5},\cite{Stieb, Kim}  
(for a review see \cite{munoz} )
the Lagrangian is extremely complicated, and many 
details remain to  be understood more thoroughly.
The present paper derives some further aspects of the 
low energy Lagrangian. Our main result is a  formula 
for the potential energy for the moduli fields, 
valid in certain regions of moduli space. 
The detailed expression is given 
in eq. (\ref{U}) et. seq. below, for the case when there is a 
single 5-brane insertion and $h^{1,1}(\cX)=1$. 
Since the derivation is rather long we  
explain here a few of the ingredients of this formula. 

The chiral scalars in $d=4$ supergravity   take 
values in a target space which is K\"ahler-Hodge. 
These   fields   correspond to moduli for the
Calabi-Yau metric on $\cX$, moduli for the instanton 
gauge field along $\cX$, and moduli for the positions 
of the M5 branes along the interval. In addition 
there are chiral fields charged under the gauge group $H$
left unbroken by the $E_8 \times E_8$ instanton. 
These will   generically be denoted by $C^I$. 

The superpotential $W$ is a section of a line bundle on 
the Kahler-Hodge target space for the chiral scalars.  
There are several sources for the superpotential in the 
effective supergravity. First of all, there is a 
perturbative term, cubic in
the scalars $C^I$. In addition, there are several 
sources of nonperturbative effects. Some of these, such as 
heterotic worldsheet instantons, gluino condensation, and 
M5 instantons (wrapping $\cX$) have been studied in 
many previous papers \cite{Witten2}-\cite{Hor}.
The inclusion of the effects of
 open membrane instantons, which have not been 
studied as thoroughly, is the main focus of this paper.

There are three kinds of open membrane effects we must consider, 
since M2 branes can end both on M5-branes, or on the boundary 
``M9 brane'' \cite{Townsend,branes ending on branes}.
 Membranes stretching between the boundary 
M9-branes are the $M$-theory versions of heterotic 
worldsheet instantons, and as such have been studied in the 
context of $(0,2)$ compactifications of heterotic string 
backgrounds \cite{Den1, Den2}.
 It is well-known that such effects often 
sum to zero, e.g., in backgrounds admitting a description 
by a linear sigma model  \cite{Silv,Kach,candelas3}.
The mechanism by which these contributions vanish is that 
a given homology class can contain many different holomorphic 
curves in $\cX$. The instanton action depends only on the 
homology class, but the prefactor depends on the curve, and 
the sum of instanton amplitudes can, and often really does,
vanish, as can already be seen in the case of the quintic. 
By contrast, the M2 instantons stretching between 
M5 and M9, or between M5 and M5 must wrap the particular 
holomorphic curve $\Sigma$ already wrapped by the M5 brane. This is 
obvious for the part of the membrane worldvolume 
ending on the 5-brane. A study of the  
conditions for the supersymmetric instanton (based on \cite{BBS})  
reveals that the membrane must have a direct 
product structure $\Sigma \times I$ where $I$ is an 
interval and $\Sigma \subset \cX$ is a holomorphic curve. 
(The detailed argument is given in section 3 below.) 
Consequently, if $\Sigma$ is a rigid holomorphic curve in 
$\cX$ there will be no sum over instantons, and 
no integral over the moduli space for the curve. 
Moreover, if $\Sigma$ is a rational curve there will be 
precisely two fermion zeromodes and the fermion 
2-point function determining the superpotential 
will be nonzero.  
(Our calculation of  the induced superpotential uses  
the technique discussed in \cite{Witten2,BBS,HM}.)

The backgrounds we study are in a regime of $M$-theory 
where we can do systematic expansions in the 
long wavelength expansion. It follows 
from  \cite{HorWit, HorWit1, Witten} that 
this is an expansion in $R/V^{2/3}$ where 
$R$ is the length of the interval $S^1/Z_2$ 
and $V$ is the volume of $\cX$ in 11-dimensional 
Planck units. We therefore assume $R/V^{2/3}\ll 1$. 
Now, gluino condensation and 5-brane instanton 
effects contribute terms of order 
$\Delta W \sim \exp[- c_1 V]$ to the 
superpotential $W$, where $c_1$ is of order 1. 
By contrast, open membrane effects contribute terms 
of order $\Delta W \sim \exp[- c_2 R V^{1/3} ] $ 
where $c_2$ is of order 1 (or smaller). 
Thus, in the backgrounds under study in this paper,
 {\it open membrane instantons 
are the leading source of nonperturbative effects}.

Our goal is to understand the physics of the moduli 
in heterotic M-theory, so we need the potential, 
rather than just the superpotential. 
The potential energy for scalars in $d=4,N=1$ 
supergravity is 
given by the famous formula \cite{cremmer,WB}
\be
{(\kappa_4)}^{4}U=e^{K}\bigl(K^{i\bj}D_i W\overline{D_jW}-3 W\bW \bigr)
 + U_D
\label{U0}
\ee
where $2\kappa_4^2=16\pi G_N$ is the (four-dimensional) 
Newton constant, $K$ is the K\"ahler potential, 
$D_iW=\partial_i W + \partial_i K W$ is the 
covariant derivative, and $U_D$ are ``D-terms'' for charged scalars 
$\sim \sum_a (\bar C T^a C)^2$. 

The potential (\ref{U0}) is extremely complicated. Moreover, $K$ is 
 only approximately  known only in some regions of moduli space. 
We are therefore forced to consider perturbation expansions 
in several quantities. First, we will expand in two 
dimensionless parameters 
\be
\ce\sim R/V^{2/3}\ll 1 \qquad \quad \cer \sim V^{1/6}/R\ll 1
\label{expparams}
\ee
which are necessary for the validity of the geometrical 
11-dimensional picture (more precise formulae 
appear in eq. (\ref{eff}) below). Note that these imply that 
$V\gg 1$ and $R\gg 1$, and that the length of the interval is 
much larger than the scale set by $\cX$. 
In addition we must expand in powers of the charged scalars 
$C^I$. The superpotential is a sum of two terms 
$W= W_{\rm pert} + W_{\rm nonpert}$, where $W_{\rm pert}$ 
is a cubic expression in the charged scalars $C^I$ 
with coefficients that are functions of the  
complex structure and bundle moduli. 
We can organize the terms according to 
whether they are order $0,1$, or $2$ in $W_{\rm nonpert}$: 
\be
(\kappa_4)^4U = (U_{0} + U_{1} + U_{2})
\label{schematic} 
\ee 

We will now describe the leading expressions for the
three 
terms in (\ref{schematic}) in 
the case of a Calabi-Yau $\cX$ with 
$h^{1,1}(\cX) = 1$ together with   a single 5-brane,
inserted at $x$, where $0\leq x \leq 1$ labels the 
position of the 5-brane along the M-theory interval. 
In addition to the charged scalars $C^I$ the
 relevant chiral superfields are the ``volume superfield'' 
$S = V+ i \sigma$, which determines the GUT coupling, 
the ``K\"ahler superfield''  $T= Ra + i \chi$, where 
$a$ is the K\"ahler modulus for  $\cX$ (hence 
$V\sim a^3$),  and 
the ``position superfield'' $Z= R a x + i \alpha $
for the 5-brane. The fields $\sigma, \chi$ and $\alpha $ are 
axions. 

 The first term, $U_0$, in (\ref{schematic}) begins with 
 the perturbative contribution 
to the potential. The leading order expression in 
an expansion in the charged scalars is 
  a positive semidefinite quartic form: 
\be
\displaystyle
U_0 = \frac{1}{V J^2} U_{I\bar J K \bar L} C^I \bar C^{\bar J} 
C^K \bar C^{\bar L} \biggl(1+ \cO(\frac{ C^2}{J}, \ce,\cer)\biggr)
\label{firstu} 
\ee
Here $J:= Ra$.
  The coefficients $U_{I\bar J K \bar L}$  are functions of 
the complex structure and 
bundle moduli. We will give precise formulae for them, but 
will not be very explicit about their behavior.

The leading contribution 
to the second term in (\ref{schematic})
is a one-instanton term resulting from cross terms 
between the perturbative and nonperturbative 
superpotentials. We find that the single instanton contribution has the
form 
\be
U_{1} =   \frac{(1-x) }{V J^2}\Bigl \{e^{-Jx}
\Re\bigl[U_{IJK} C^I C^J C^K e^{i \alpha}\bigr] -  
 e^{-J(1-x)} \Re\bigl[U_{IJK} C^I C^J C^K e^{i (\chi - \alpha)}\bigr]
\Bigr \}
  +   \cdots 
\label{schematicii}
\ee
 The coefficients $U_{IJK}$ are 
functions only of the complex structure and 
bundle moduli.

Finally, the third term $U_2$ in (\ref{schematic})
begins with a 2-instanton effect
\be
\displaystyle
U_2 = \frac{E}{J^2}  \Biggl \{
 e^{-2Jx} + e^{-2J(1-x)} - 2e^{-J} \cos(2\alpha - \chi)
\label{schematiciii}
\ee
$$+ \frac{2J}{3V}\bigl (1-2x\bigr )e^{-2J(1-x)}+
\frac{4Jx}{3V}e^{-J}cos\bigl (2\alpha - \chi \bigr )
  +\cdots 
 \Biggr \}
\biggl(1+ \cO(\frac{ C^2}{J}, \ce,\cer)\biggr)$$
where $E$ is a positive definite function that depends only on 
the complex structure and bundle moduli. (We have kept some 
subleading terms in the second line. The reason for this 
is explained in detail in sections 5.4 and 5.5.) 

A precise characterization of the region of validity of the 
above potential is 
given in section 5.4 below. The strongest constraints 
on the region of validity come from our ignorance of 
the exact K\"ahler potential. It is also important to 
bear in mind that the coefficients of the 
higher order terms in the expansion in 
$\frac{\vert C \vert^2}{J}, \ce, \cer$ are functions of the complex structure and bundle 
moduli. If these coefficients become singular somewhere 
in the moduli space then these ``higher order'' terms 
will dominate the physics. Our working assumption is 
that we are at a generic smooth point in bundle and complex 
structure moduli space.

Having determined the leading nonperturbative effects, 
and thereby the potential energy, we
investigate briefly some of the resulting classical 
dynamics on moduli space, at a somewhat heuristic level.  
Although the M5 branes wrap all of spacetime, thanks to 
the central term in the superalgebra, their positions 
along the M-theory interval
are in fact dynamical variables.  In the regions where we can 
trust our answer  we find two kinds of instabilities in 
the compactification, depending on whether the effects 
of vevs of the charged scalars $C^I$
are important or not.
 When the charged scalar vevs are 
important, the leading $x$-dependent effect 
is a one-instanton effect. The axions will evolve to 
produce an attractive force between the 
M5 brane and the nearest M9-wall. This could possibly 
be interpreted as a consequence of the 
Witten effect: the axions evolve and continuously 
change an effective brane charge in order 
to produce ``the most attractive channel,'' in particular 
producing an attraction between the 5-brane and 
the boundary.  It would 
be interesting to understand the physics of this 
effect more fully.  

The above discussion is valid for $Jx \gg 1.$ 
As the five-brane moves towards the wall the approximations 
break down. The limit $x\to 0$ is extremely interesting 
and is related to the chirality-changing transitions 
discussed in   \cite{Kachru2, smallinst}. In order to 
study this limit one needs a multiple cover formula for 
the membrane instantons. This is discussed in section six
below. We make some educated guesses and conclude that 
the physics depends on the (unknown) details of the covering formula.

A second kind of instability occurs when charged scalar 
vevs are small or zero.  In this case the potential 
has a local minimum in $x$ at $x=\half.$
The value of $U$ at such points is small and of the form   
$$\displaystyle
U \sim \gamma \frac{ e^{- J}}{J V} $$
where    $\gamma$   is a positive function
of the complex structure and bundle moduli.
The M2 branes lead to a repulsive 
interaction between the M5-brane and the M9-brane 
which induces decompactification of both the M-theory 
radius and the Calabi-Yau, while the M5-brane
moves to the middle of the interval. Of course, in this 
instability new light modes appear as the theory becomes 
five-dimensional, and we should describe a matching to 
a description in terms of five-dimensional supergravity. 
(As the M5 moves to the middle of the interval there 
is a balancing of forces from the two boundaries 
and  the 
leading terms in $U_2$ vanish. This is why we must 
include the subleading terms.)

The second kind of  instability is an 11-dimensional manifestation of 
the Dine-Seiberg problem; it is hardly unexpected, and in 
the case of the standard embedding similar instabilities 
have already been pointed out by Banks and Dine in \cite{Dine}. 
Nevertheless, it is interesting to note that in the 10-dimensional 
Dine-Seiberg instability the size of the M-theory interval 
$S^1/Z_2$ tends to {\it shrink}. There are thus different 
asymptotic regions of moduli space with qualitatively 
different dynamics, and hence different ``basins of attraction'' 
for the classical evolution of the moduli. One consequence is 
that there must be nontrivial stationary points for the 
potential in the middle of moduli space. The precise nature of 
such stationary points is of great interest, but remains out 
of reach so long as we cannot derive the K\"ahler potential 
in the interior of moduli space in a controlled approximation.

The paper is organized as follows.
In     section two we review briefly the 
$M$-theory geometry corresponding to 
strongly coupled $E_8 \times E_8$ 
heterotic strings with 
``nonstandard embedding.'' 
In section  three we study supersymmetric 
 M2-brane instantons in $\cX \times S^1/Z_2.$ 
In section four we derive the formula for the 
contribution to the superpotential from M2 instantons. 
In   section five 
we find the  potential and specify the
 region where we can trust it for the simplest case
 of a Calabi-Yau with  $h^{(1,1)}=1.$
In section six we discuss 
the multiple covering formula and its relevance 
to chirality-changing transitions. 
 In section seven we generalize the result 
to the case of N 5-branes
on the interval. The final section contains a  discussion of 
some possible extensions of the present work.

We have been informed that the 
 effects of open membranes in heterotic M-theory 
are also being investigated independently by 
B.Ovrut, E.Lima and R.Reinbacher.

\section{Review of heterotic M-theory
 background with  M5-branes on the interval}

In this section we review some of the results of 
 (\cite{Witten,Dine},\cite{Ovrut1} - \cite{Ovrut5},\cite{munoz}) which 
are needed for our subsequent computations.  

Our conventions for  the Lagrangian of 11D SUGRA 
are set by the Lagrangian: 
\be
2\k S_{11D}=-\int eR -\half \int G_4\wdg*G_4 
-\frac{1}{6}\int C_3\wdg G_4\wdg G_4
+ \ldots
\label{SUGRA}
\ee
where $G_{MNPQ}=4\partial_{[M}C_{NPQ]}$.
\footnote{We have  a different normalization
of fields compared to (\cite{Ovrut1}).
 $G_{MNPQ}^{here}={\sqrt{2}}G_{MNPQ}^{\cite{Ovrut1}},$ 
 $C_{MNP}^{here}=6{\sqrt{2}}C_{MNP}^{\cite{Ovrut1}}.$ 
We use the convention $2\k=(2 \pi)^8 {(M_{11})}^{-9}$, and 
define the 11-dimensional Planck length by 
$l_{11}=1/M_{11}$.  Our signature
is mostly plus.}

The Lagrangian of the boundary $E8 \times E8$
theory is given by
\be
\displaystyle
2 \k S_{YM}=-\frac{1}{4 \pi}\Bigl (\frac{\kappa_{11}}{4 \pi}
\Bigr )^{\frac{2}{3} }\int_{M_1^{10}}\sqrt{-g}tr\Bigl (F^{(1)} \Bigr )^2
-\frac{1}{4 \pi}\Bigl (\frac{\kappa_{11}}{4 \pi}
\Bigr )^{\frac{2}{3} }\int_{M_2^{10}}\sqrt{-g}tr\Bigl (F^{(2)} \Bigr )^2
\label{boundary}
\ee
where $F^{(1,2)}$ are the field strengths of the two $E_8$ gauge fields, 
to leading order in a long-wavelength expansion. In the above action and
below
$tr$ means $\frac{1}{30}$ of the trace in the adjoint of $E_8.$

We begin by describing the background
solution of $M$-theory on $R^4\times \cX \times S^1/Z_2$. 
Our coordinates on 
$R^4$ are $x^\mu$,  $\mu=1,\ldots,4$.
Complex coordinates along $\cX$ have indices $m,\bm=1,\ldots,3$.
The factor $S^1/Z_2$ in  spacetime has coordinate $X^{11}$. 
In addition it 
will be convenient to set 
$X^{11}=\pi\r y $ where $y$ is a dimensionless coordinate 
$0\leq y\leq 1$, and $\r$ is a dimensionful constant 
which sets a scale.

We must now specify the 
metric, four-form $G_4$, 
and boundary Yang-Mills fields. 
In order to write the background metric   we introduce a basis of
 harmonic (1,1) forms on $\cX$, $\omega_i, i=1,\ldots, h^{1,1}$  and 
 denote the Kahler form on  $\cX$
by    $\omega=a^i\omega_i$.  Then, the background metric is 
a deformation of a metric 
of  the   form 
\be
\displaystyle
ds_{11}^2=V^{-1}{R}^{-1} g_{\mu\nu}dx^{\mu}dx^{\nu}+
{R}^2 {(dX^{11})}^2
-2i\omega_{m\bm}dx^m dx^{\bm}.
\label{ovrutii}
\ee
In this formula $R$ is dimensionless and 
$R\r$ is the orbifold radius. Similarly, we introduce
a fiducial, dimensionful, volume $v$ for $\cX$, 
and the volume of $\cX$ in the metric (\ref{ovrutii}) 
defines the dimensionless 
parameter $V$ by $Vv := {1\over 3!} \int_{\cX}\omega^3$.  
We will make a convenient choice of $\r,v$ in 
eq. (\ref{choice}) below; 
they will be of the order of $l_{11},l_{11}^6$
and are independent of moduli. Because of the 
Weyl-rescaling in the first term in 
(\ref{ovrutii}), 
 $g_{\mu\nu}$ is the four-dimensional Einstein metric
and the four-dimensional Newton constant is given by 
\be
\frac{1}{\kappa_{4}^2}=\frac{2\pi \r v}{\k}.
\ee

As we have mentioned, the actual metric we will use is a 
deformation of eq. (\ref{ovrutii}), and is only 
known to first order in  a power series in two 
dimensionless expansion parameters
\be
\displaystyle \ce=\frac{\e R}{V^{\frac{2}{3}}}\ll 1, \quad 
\cer=\frac{\er V^{\frac{1}{6}}}{R} \ll 1
\label{eff}
\ee
where we choose the constants  
\be
\displaystyle
\e=\Bigl(\frac{\kappa_{11}}{4\pi}\Bigr)^{\frac{2}{3}}\,\,
\frac{2\pi^2\r}{v^{\frac{2}{3}}}=2,\qquad
\er=\frac{v^{\frac{1}{6}}}{\pi\r}=\bigl (\frac{\pi}{2}\bigr )^{\half}
\label{choice}
\ee
in order
to simplify the normalization of the fields in the effective 
Lagrangian.\footnote{ We take $v=8{\pi}^5 l_{11}^6, \quad
\pi \r =2 \pi^{\frac{1}{3}}l_{11}$
 to have $\e =2,\er^2 =\half \pi$}

The above inequalities (\ref{eff})  state, firstly,
that the distortion of the background
from (\ref{ovrutii}) is small, and secondly
that the interval is much larger 
than the length scale of $\cX$. 
These expansion parameters can be related to  the GUT scale and the 
4-dimensional Newton constant \cite{Witten,Dine}.
In our conventions the unified coupling $\alpha_{\rm GUT} 
\sim (\ce\cer)^2\sim 1/V $, while $(M_{\rm GUT}\kappa_4)^2
\sim (\ce)^3(\cer)^4 \sim 1/(R V^{4/3})$ determines the GUT 
scale in terms of the Newton constant. The latter formula 
follows by computing masses of gauge bosons and scalars 
associated with typical mechanisms of spontaneous symmetry 
breaking. 
\footnote{We thank T. Banks for very helpful discussions on 
this point.} 
Unfortunately, 
it turns out that when we use the experimentally 
measured values of $\alpha_{\rm GUT}, M_{\rm GUT}$
and $\kappa_4$ the above expansion is not necessarily 
a good   approximation. As discussed in
\cite{Witten,Dine,Dine2,Ovrut1,Benakli}, the experimentally 
measured values determine   $\cer \ll \ce=O(1).$
Nevertheless, our focus in this paper is on a systematic 
and controlled computation of nonperturbative effects; 
the restriction  (\ref{eff}) is necessary since 
heterotic M-theory  is only known
as an effective theory to order ${(\kappa_{11})}^{\frac{2}{3}}$, 
and for this reason we will adopt it.

To lowest order in the expansion parameter the metric for the 
background takes the form
\be
\displaystyle
ds_{11}^2=V^{-1}{R}^{-1}(1+\frac{B}{6})g_{\mu\nu}dx^{\mu}dx^{\nu}+
{R}^2(1-\frac{2B}{3}){(dx^{11})}^2
-2iJ_{m\bm}dx^m dx^{\bm},
\label{ovrut1}
\ee
$$J_{m\bm}=\omega_{m\bm}+\bigl(B_{m\bm}-\frac{1}{3}\omega_{m\bm}B\bigr),
\qquad B=2\omega^{m\bm}B_{m\bm},$$
The  deformation of the background is described by the (1,1) form
$B_{n\bar m}$. In order to write it explicitly we must now 
introduce the M5 branes.

The backgrounds we study preserve $N=1$ supersymmetry.
Therefore the 5-branes wrap a product of spacetime 
and a holomorphic curve in $\cX$. If there are $N$ 
5-branes they will therefore have definite locations at 
$y=x_k$, $k=1,\ldots, N$ along the interval. 
The $k^{th}$  5-brane wraps a curve $\Sigma^{(k)}$ in $\cX$
whose homology class may be expressed as 
$[\Sigma^{(k)}]=\b_i^{(k)}[\Sigma^i_2]$
where  $ [\Sigma^i_2] $ is an integral basis of $H_2(X,\bf{Z})$, and 
$\b_i^{(k)}$ is a collection of nonnegative integers. 
These integers are constrained by anomaly cancellation. 
Each of the M9 branes carries an $E_8$ vector bundle
$V_1,V_2$, and to each bundle  we associate a degree four integral
characteristic class $c_2(V_i)$. Identifying 
 $H_2(\cX;Z)$ with $H^4(\cX;Z)$ via Poincar\'e 
duality we may define
\be
c_2(V_1)-\half c_2(TX)=\b_i^{(0)}[\Sigma^i_2 ]\qquad
c_2(V_2)-\half c_2(TX)=\b_i^{(N+1)}[\Sigma^i_2] \qquad
\label{beta}
\ee
The   anomaly cancellation condition is then
\be
\sum_{n=0}^{N+1}\b_i^{(n)}=0.
\label{anomaly}
\ee
In terms of the above data, the formula for 
$B_{m\bar m}$    on
the interval $(x_n,x_{n+1}), \quad n=0,\ldots, N$ 
is given by
\be
\displaystyle
B_{m\bm}=\frac{2R}{V} b_i \omega^i_{m\bm},\qquad
b_j(y)=\sum_{k=0}^{n}\b_j^{(k)}(y-x_k)-
\half\xi_j, \qquad \xi_j= \sum_{k=0}^{N+1}(1-x_k)^2\b_j^{(k)},
\label{b_j}
\ee
where $x_0=0, x_{N+1}=1$ and 
 the  index $i$ is raised with the inverse of the metric on the
 moduli space of Kahler structures on $\cX$:
\be
G_{ij}=\frac{1}{2vV}\int_{\cX} \omega_i\wdg \bigl(*\omega_j
\bigr)=-\half\partial_i\partial_j
ln\Bigl(d_{i_1i_2i_3}a^{i_1}a^{i_2}a^{i_3}\Bigr)
\label{Gij}
\ee
with 
\be
d_{i_1i_2i_3}=\int_{\cX}\omega_{i_1} \wdg \omega_{i_2} \wdg \omega_{i_3}.
\label{d}
\ee
The choice of integration constant
in the solution (\ref{b_j}) fixes $Vv$ to be equal to the  volume
of $\cX$  averaged along the interval (to lowest order in 
$\ce$).

The flux of the 4-form $G_4$ is also given in terms of $B$: 
\be
G_{MNPQ}=\frac{1}{2}\epsilon_{MNPQEF}\partial_{11}B^{EF}
\label{ovrut2}
\ee
Note that it is discontinuous across the positions of the 5-branes.

Finally, we need to specify the $E_8$ gauge bundles $V_1$ and $V_2$. 
For simplicity we will follow \cite{Ovrut2}
and take the bundle $V_2$ at $y=1$ to be the trivial bundle. Accordingly, 
there is a ``hidden sector'' at $y=1$ with unbroken $E_8$ gauge group. 
The bundle $V_1$ at $y=0$ has an instanton whose holonomy lives in 
a subgroup $G\subset E_8$. The unbroken gauge symmetry is the 
commutant $H$ of $G$ in $E_8$. It is straightforward to extend 
our formulae to the case when both
$V_1$ and $V_2$ are
 non-trivial  bundles.

When we compactify M-theory on the above background, the 
physics at distances large compared to the M-theory interval 
is described by an effective $d=4, N=1$ supergravity theory. 
We now list   the massless fields   corresponding to small fluctuations 
around the above background. In addition to the superYang-Mills
and supergravity multiplets there are a number of massless 
chiral scalar fields. To begin with, there
  are chiral superfields neutral under four dimensional gauge group
$H$. These are: 
\be
 T^i=Ra^i+i \chi^i, 
\label{tee}
\ee
\be
  S=V+i\sigma,
\label{ess}
\ee
\be
Z_n=R(\b_i^{(n)}a^i)x_n-i\Bigl[{\cA}_n(\b_i^{(n)}a^i)-x_n(\b_i^{(n)}\chi^i)
\Bigr ]
\label{list}
\ee
where
$$C_{m\bm11}=\chi^i\omega_{i, m\bm}, \qquad i=1,\ldots,h^{1,1}, \quad
m,\bm=1,\ldots,3, $$
$\sigma$ is a scalar dual to $C_{\mu\nu11}$
$$3\partial_{[\mu}C_{\nu \r ] 11}=
V^{-2}\epsilon_{\mu \nu \r \lam}\partial^{\lam}\sigma,$$
and  $Z_n$ is a holomorphic coordinate
constructed out of
the position $x_n$  of the n-th 5-brane on the interval.
The scalar ${\cA}_n$ originates 
from the KK reduction of the 2-form living on the 
n-th 5-brane  
\be
\displaystyle 
A^{(2)}_n=  \pi \r  {\cA}_n f_n^*(\omega) 
\label{defA}
\ee
We have included the factor $\pi \r $ in the above formula to make
 $ {\cA}_n $
dimensionless.
 In   eq.(\ref{defA})
$f_n^*(\omega)$ is the pullback of the Kahler form to the 
cycle $\Sigma_2^{(n)}.$
We denote   by $f_n $ the holomorphic embedding of the curve
 $\Sigma_2^{(n)}$ in $\cX.$
The pullback of each of the basis forms $f_n^*(\omega_i)$
 is proportional
to the pullback of the Kahler form $\omega$
$$\displaystyle 
f_n^*(\omega_i)=\frac{\b_i^{(n)}}{(\b_j^{(n)}a^j)}f_n^*(\omega), \quad
\int_{\Sigma_2^{(n)}}f_n^*(\omega_i)=v^{\frac{1}{3}}\b_i^{(n)}.$$

Finally, there are chiral multiplets charged under 
the unbroken gauge group $H$.  Thanks to the 
Donaldson-Uhlenbeck-Yau theorem
massless modes from small fluctuations 
of the gauge field can be associated with 
 holomorphic deformations of 
  holomorphic bundles on $\cX$.  The small fluctuations are
parametrized 
by the space $H^{0,1}_{\bar\partial}(X, V )$
where $V$ is the gauge bundle in the ${\bf 248}$. 
We assume the holonomy of the instanton is in $G$ so
the gauge bundle decomposes as 
$V = \oplus W_{\cal R} \otimes V_{\cal S}$ 
corresponding to the decomposition of 
the adjoint of $E_8$ under  the 
embedding $H\times G\subset E_8$:
\be
\mathbf{ 248 =\oplus {\cR}\otimes {\cS}}
\label{248}
\ee
  The charged scalars will be valued in 
$\oplus W_{\cal R}\otimes H^{0,1}_{\bar\partial}(X, V_{\cal S} )   $. 
In order to work out the Kaluza-Klein reduction we  
 decompose the gauge field as: 
\be
\displaystyle
A_{\bm}=\frac{2^{3\over{2}} \pi}{\kappa_4 }
u_{\hI, \bm}C^{\hI}, \quad \bm =1, 2, 3,
\label{gauge}
\ee

 In (\ref{gauge}) a  summation is taken over 
 the index $\hI$ which labels
$$\hI=(\cR,I,p), \quad p=1, \ldots, \dim\cR,\quad
I=1,\ldots \dim H^1(X,V_{1 \cS }).$$
The normalization factor in (\ref{gauge}) was chosen 
to make the charged scalar fields $C^{\hI}$ dimensionless and to
normalize their  kinetic term conveniently. 

When writing the perturbative superpotential below it will 
be convenient to define 
$$u_{\hI, \bm}=u^x_{I\bm} T_{xp} ,$$ 
where $x$ is an index for a basis for the 
representation $\cS$ and 
 $ u^x_{I\bm}$ is a basis of $H^1(X,V_{1 \cS }).$
The factor 
 $T_{xp}$ is purely group-theoretic and 
corresponds to the generators of $E_8$ in 
the representation  $\cR \otimes \cS$. The complex conjugate
of these generators is denoted by  $T^{xp}$ and the normalization is
chosen such that
$tr\Bigl (T_{xp}T^{yq}\Bigr )=\delta_x^y \delta_p^q .$

We are not going to study four-dimensional gauge dynamics 
in this paper. This has been studied,
 for example, in \cite{Witten, Dine, Ovrut2}.
For completeness, and to fix our normalizations, we also give
the gauge kinetic term in the 4D Lagrangian
\be
\displaystyle
S_{YM}=-\sum_{\a=1}^{2}
\frac{1}{64\pi^2 }
\int_{M^{(\a)}_4} \sqrt{-g_4}
\Bigl (Ref^{\a} trF^2+\ldots
\label{4dgauge}
\ee
where  due to the restrictions (\ref{eff})
on the moduli space, $Ref^{\a}=V+ O(\ce), \quad \a=1,2.$

\section{ M2-brane instantons in $\cX\times S^1/Z_2$}

Open M2-branes ending on an M5 brane 
will play a crucial role in our
calculation of the non-perturbative potential.
These nonperturbative effects were 
first discussed in \cite{Townsend,branes ending on branes,Brax}.
In this section we will derive the  conditions
for a supersymmetric open M2-brane  instanton in the background 
described in the previous section. 
 We will neglect the distortion
of the background metric from a direct product metric
in solving for the membrane configuration. This is valid in 
our approximation scheme. 
 
The first step in finding the supersymmetric M2 configuration 
is to write the constant spinors corresponding to the supersymmetries 
unbroken by the background. 
We use   a basis for the $\Gamma$-matrices in eleven 
dimensions of the form
\bear
\displaystyle
{\Gamma}^{\mu}=(RV)^{\half}\gamma^{\mu}\otimes{\gamma }^7 &
 \Gamma^m=1\otimes\gamma^m&
\mu=1,\ldots,4, \quad \{\gamma^{\mu},\ \gamma^{\nu} \}=2 g^{\mu \nu}\\
\Gamma^{\bm}=1\otimes\gamma^{\bm}& 
\Gamma^{11}=\frac{1}{R}{\gamma }^5 
\otimes{\gamma }^7 & 
m,\bm=1,\ldots,3, \quad \{\gamma^n, \ \gamma{^\bm} \}=2g_{CY}^{n \bm}
\eear
where $(\gamma_m)^*=-\gamma_{\bm}=(\gamma_m)^T$
and $\gamma^{\mu}$ is a weyl-basis in 4D.

Four dimensional anti-chiral ( chiral) spinor 
indices are denoted by $\a$ ($\dot \a$) respectively. 
In this basis the  surviving supersymmetry in the background
$\cX \times S^1/Z_2$
is of the form:
\be
{\bf \epsilon}=\Bigl (\epsilon^{\dot \a}\otimes\epsilon_1, 
\epsilon^{ \a}\otimes\epsilon_2\Bigr ),
\label{eps}
\ee
where $\epsilon^{\dot \a}, \epsilon^{ \a}$ 
are constant spinors on $R^4 \times S^1/Z_2$ and 
$\epsilon_1$ ($\epsilon_2=(\e_1)^*$ ) is the chiral( anti-chiral)
 covariantly constant spinor on $\cX$, 
 normalized as in \cite{BBS}:
\be
\gamma_{\bm}\epsilon_1=0, \quad \gamma_{n\bm
}\epsilon_1=i\omega_{n\bm}\e_1,
 \quad \gamma_{mnp}\e_1=e^{-K}\Omega_{mnp}\e_2,\quad \e_1^\dagger \e_1=1 .
\label{norma}
\ee

Here $\omega$ is the Kahler form, $\Omega$ is a holomorphic
(3,0) form on $\cX$ and $K=\half (K_T-K_{cplx})$
with both Kahler functions $K_T$ and $K_{cplx}$ specified in
section (5.2).

The surviving supersymmetry is consistent with having 5-branes 
wrapped over a holomorphic
cycle $\Sigma \subset \cX,$ as   shown in \cite{Ovrut1}.
  One cannot have anti-5-branes
 on the interval and preserve supersymmetry. 
 
 The presence of an M2-brane imposes an additional constraint
 on the supersymmetry
 parameter ${\bf \epsilon}$
\be
\Gamma^{(2)}{\bf \epsilon}={\bf \epsilon},
\label{22}
\ee
where, (see for example \cite{BBS}), 
\be
\displaystyle
\Gamma^{(2)}=\frac{i}{3!\sqrt{g}}\e^{ijk}\partial_{i}X^{\hat M}
\partial_{j}X^{\hat N}
\partial_{k}X^{\hat K}
\Gamma_{{\hat M}{\hat N}{\hat K}}
\label{gamma2}
\ee
In formula (\ref{gamma2}) $s^i, i=1,2,3$
 are coordinates on the world-volume of the M2-brane, $X^{\hat M}, \ 
{\hat M}=(\mu, m ,\bm, 11)$
are coordinates in the eleven dimensional target space and
$g$ is the determinant of the induced metric on the M2-brane.

Substituting (\ref{eps}) into (\ref{22}) we find, first of all, that 
spinors of type 
$ \epsilon^{ \a}\otimes\epsilon_2$
lead to  
\be
\displaystyle
\epsilon_2=
\Bigl (
\frac{R}{\sqrt{g}}\e^{ijk}\partial_{i}X^m\partial_{j}X^{\bn}
\partial_{k}X^{11}\omega_{m \bn}\Bigr )\epsilon_2+
\Bigl (\frac{ie^{-K}}{3!\sqrt{g}}\e^{ijk}\partial_{i}X^{\bm}
\partial_{j}X^{\bn}\partial_{k}X^{\bp}
\Omega_{{\bm}{\bn}{\bp}} \Bigr ) \epsilon_1
\label{gam2e2}
\ee
$$+\Bigl (\frac{\e^{ijk}}{\sqrt{g}}\partial_{i}X^m\partial_{j}X^{\bn}
\partial_{k}X^{\bp}\omega_{m \bn}\Bigr )\gamma_{\bp}\epsilon_2+
\Bigl (\frac{ie^{-K}}{4\sqrt{g}}\e^{ijk}\partial_{i}X^{\bm}
\partial_{j}X^{\bn}\partial_{k}X^{11}
\Omega_{{\bm}{\bn}{\bp}}g_{CY}^{\bp q} \Bigr )\gamma_{q} \epsilon_1
$$

Since the spinors $\e_1,\e_2,\gamma_{m}\epsilon_1,\gamma_{\bm}\epsilon_2$
 are linearly independent  we get four equations  
\be
\partial_{i}X^{\bm}\partial_{j}X^{\bn}\partial_{k}X^{\bp}
 \Omega_{{\bm}{\bn}{\bp}}=0
\label{eq1}
\ee 
\be
R\partial_{i}X^m\partial_{j}X^{\bn}
\partial_{k}X^{11}\omega_{m \bn}=\sqrt{g}\e_{ijk}
\label{calibration}
\ee
\be
\partial_{i}X^{\bm}\partial_{j}X^{\bn}\partial_{k}X^{11}
 \Omega_{{\bm}{\bn}{\bp}}g_{CY}^{\bp q}=0
\label{eq2}
\ee 
\be
\e^{ijk}\partial_{i}X^m\partial_{j}X^{\bn}
\partial_{k}X^{\bp}\omega_{m \bn}=0
\label{special}
\ee
The constraints (\ref{eq1}, \ref{special}) are automatically
solved by the embedding

$$X^{11}=t, \quad X^m(y)$$
where $t=s^3$ is a coordinate along the orbifold interval
and $y, \by$ are  coordinates on a holomorphic 2-cycle.
This is our basic instanton. 

We claim that if the holomorphic curve 
$\Sigma \subset \cX$ is isolated then the above 
membrane instanton is also. Moreover, we claim 
that the above instanton is the only instanton 
solution consistent with the boundary condition 
of having the M2 brane ending on $\Sigma$.  Indeed, let us consider  
 the possibility of  having M2-branes 
starting and ending on 
holomorphic cycles inside $\cX$ which differ   from a direct product
$\Sigma \times I.$
Therefore we search for   $t$-dependent 
embeddings $X^m(y,t), t\in[x_1,x_2]$ into   $\cX.$
In this case equation (\ref{special}) is not satisfied automatically
and gives the constraint
\be
\partial_{[i}X^m\partial_{j]}X^{\bn}\omega_{m \bn}=0,
\label{imp}
\ee
Taking the  $i=y, j=\by$ component of this equation and evaluating  it at
the boundary
$t=x_1$ or $t=x_2$ shows  that 
  the volume of the holomorphic
cycle must be zero. 

We conclude that  an open M2-brane which
starts and ends on a positive volume  holomorphic curve 
preserves some supersymmetry iff it has 
the direct product form $\Sigma\times I.$

One can quite analogously prove that
 an M2-brane which starts
and ends on a holomorphic curve should have the direct
product form  
$$X^{11}=-t, \quad X^m(y)$$
 in order to preserve the other components  
  $ \epsilon^{\dot \a}\otimes\epsilon_1$
of the background supersymmetry.

 Note that since the M2-brane instanton must
 start and end on the same 2-cycle in $\cX$
 there is a requirement on the
5-brane charges $\b_i^{(n)}=\b_i^{(k)}$ described in section 2
in order for there to be an   M2-instanton   stretched 
 in the interval  $[x_n,x_{k}].$

 \section{Calculation of membrane-instanton-induced superpotentials}

In this section we will give the derivation of the non-perturbative
four-dimensional superpotential $\Delta W$ induced by open membranes.

We follow the procedure outlined in \cite{BBS,HM}. 
The idea is to compute the 2-point correlation function of
four-dimensional
fermions with the instanton sector included in the supergravity 
path integral. An essential ingredient of this calcultaion is the
 coupling of the four-dimensional fermions to the world-volume
 degrees of freedom of the membrane through the  so-called 
``membrane vertex operators.'' The computation of the 
superpotential follows from a computation of a 2-point correlation  
function of fermions in the four-dimensional effective theory
 $\langle \chi\chi\rangle_{inst}$, where $\chi$ are   fermionic
superpartners of $Z$. 
This in turn can be reduced to a membrane path integral with
corresponding
 vertex operator insertions.

\subsection{Summary of the computation of $\Delta W$} 

Since the analysis is rather long let us summarize the 
computation here. Most of the work is devoted to finding the vertex
operator, 
but the end result is very simple. The membrane theory has a chiral 
doublet of fermions $\vth^{\dot \a}$ transforming in the ${\bf 2}$ 
of the 4 dimensional Lorentz group. These couple to the chiral 
fermions $\chi_{\dot\a}$ in the superfield $Z$ via the vertex operator 
\be
\displaystyle
V_{\chi}=\frac{i}{2}\vth^{\dot \a }\chi_{\dot \a}.
\label{coupl}
\ee

Using the above coupling
 we can compute $\langle  \chi(\xi_1)  \chi(\xi_2)\rangle $ in 
an instanton sector to be
\be
\displaystyle
 \int \sqrt{-g_4} d^4\xi   S_{F}(\xi_1-\xi)S_{F}(\xi_2-\xi)h \Phi
exp\bigl( - Z\bigr ).
\label{cor}
\ee
Here $\xi_1,\xi_2$ are points in four dimensions
and $S_F$ is the 4-dimensional  fermion propagator
in the effective $d=4,N=1$ supergravity. This expression for 
the propagator is only valid for $(\xi_1-\xi), (\xi_1-\xi_2), (\xi_2-\xi)
\gg l_{11}$. 
The integral of $\xi^\mu$ in   eq.(\ref{cor})
should be regarded as an integral over 
the bosonic zero modes $X^{\mu}=\xi^{\mu}$
of the M2-instanton. The integral over  the 2 fermion zero-modes
 $\vth^1,\vth^2,$  on  the M2-brane soaks up the $\vartheta^{\dot\alpha}$ 
from the vertex operator. There are no 
other zero modes because the curve $\Sigma$ is a rational 
curve and hence has   no
extra zero-modes associated with 1-forms.
The prefactor $h \Phi$ stands for determinants of fluctuations in 
11-dimensional supergravity together with 5-branes around the 
background
(\ref{ovrut1},\ref{ovrut2}), together with determinants 
associated with  the degrees of freedom for the M2 instanton. 
While it is very complicated one can use holomorphy to extract the 
factor $h e^{-Z}$, which depends holomorphically on the moduli. 
The factor $h$ is   a   holomorphic section of a
line bundle over  complex structure moduli space and 
should properly be regarded as the true measure for the 
fermion zeromodes. In this 
paper we will not be very explicit about it.

We can now extract $\Delta W$  by comparing (\ref{cor}) 
with the 2-point correlation function in the effective 4D supergravity
\be
\langle  \chi(\xi_1)  \chi(\xi_2) 
 exp[\int \sqrt{g_4} e^{\half K} \partial_Z\partial_Z(\Delta W) \bar\chi \bar \chi]
\rangle_{4D}
\label{compare}
\ee

which is equal to
\be
\int\sqrt{g_4} d^4 \xi S_{F}(\xi_1-\xi)S_{F}(\xi_2-\xi)  e^{\half K_0} 
\partial_Z\partial_Z(\Delta W)
\label{compare2}
\ee
where $K_0=K_T +K_S+K_{cplx} +K_{bundle}$ and we
drop corrections of the order $\displaystyle
O\Bigl (\ce,\cer, \frac{|C|^2}{Ra} \Bigr )$
to the mass term for a chiral fermion 
in the 4D, N=1 Lagrangian  \cite{cremmer,WB}.

Using  holomorphy of the superpotential it 
now follows that  
\be
\displaystyle
\Delta W=  h  exp\bigl( -Z\bigr), \quad \Phi= e^{\half K_0}.
\label{W}
\ee
 For the M2-brane stretched between the 5-brane and the other
9-brane at $y=1$ analogous considerations give
\be
\displaystyle
\Delta W= h exp\bigl( -(\b_iT^i-Z)\bigr).
\label{W1}
\ee

\subsection{Computation of the vertex operator} 

In this section we describe 
the computation of the  vertex operator.

The vertex operators can be found by expanding the action of the $M2$ 
brane in the M-theory background fields. 
The action of an M2-brane ending on an M5-brane
was written in \cite{Sezgin},
 using the superembedding approach of \cite{Szg&Howe,  Sorokin}.
In this approach the basic ingredients are: 

\begin{itemize}
\item{ An 
$(11|32)$ supermanifold  $M$, giving  the
11-dimensional supergravity background. The supercoordinates
are denoted  by $Z^{\underline{M}}=(X^{\hat M},\Theta ^{{\hat \r}})$.
where
${\hat \r}$ is an index in the  irreducible spinor representation of
$so(1,10).$
Using the torsion constraints of \cite{11tor}
on the supervielbein one can expand an orthonormal frame 
for the cotangent space  in  powers of $\Theta$. 
The expansion at low orders 
 in $\Theta$ has been  worked out in \cite{HM,shibusa, plefka}.
 
It is convenient 
to introduce the notation for the vielbein: 
\be
E^{\underline A}(Z)=dZ^{\underline M}E_{\underline M}\ ^{\underline A}=
(E^{\underline a}, E^{\underline\alpha})
\label{elfbein1}
\ee
 where in the second equality we have separated bosonic and 
fermionic cotangent vectors. }

\item{A $(6|16)$ supermanifold ${\cal M}$ describing the world-volume of 
the M5-brane. We denote   supercoordinates on the worldvolume 
by  $z^M=(y^M,\vth ^{\r})$ and a  cotangent frame on ${\cal M}$ 
by $e^A(z)$. A decomposition of the frame 
analogous to (\ref{elfbein1}) is given by 
\be
\label{6bein}
e^A(z)=dz^Me_M \ ^A=(e^a,~e^{\b q})
\ee
 The 
index $a=0,1,...,5$ is the index of the vector representation of 
$so(1,5)$, while  $\b$ and $q$ are the indices of 
irreducible spinor representations
of $so(1,5)$ and $so(5)$, respectively.
}

\item{A $(3|0)$ manifold $\Sigma$, to be identified with 
the membrane worldvolume. The boundaries of $\Sigma$
lie inside ${\cal M}$ or in $\partial M.$
 We denote the coordinates on $\Sigma$ by  $s^i, i,\ldots, 3$. 
Coordinates on the boundary surface are denoted by $ \sigma^r, r=1,2.$ 
}
\end{itemize}

The relation of the pullback of the 11-dimensional supervielbein 
to the 5-brane to the supervielbein of the 5-brane itself 
defines the   ``embedding matrices''  
$E_{A} \ ^{\underline{A}}$ via the equation 
\be
f^*(E^{\underline{A}})=e^A E_{A} \ ^{\underline{A}}
\label{pullback}
\ee
One may solve for these   matrices in terms of the vielbeins  
\be
E_{A}^{~~ \underline{A}} =e_A \ ^M \partial _M Z^{\underline{M}}
 E_{\underline{M}} \ ^{\underline{A}}.
\label{embmat}
\ee
 The basic superembedding condition then says that 
\be
E_{\rm fermionic}^{\qquad \rm bosonic}=
E_{\b q} \ ^a=0.
\label{supemb}
\ee
This simple equation is extremely powerful, it leads to a complete 
set of    covariant equations of motion  for the 5-brane 
\cite{Szg&Howe,Sorokin}.

The action of an M2-brane ending on an M5-brane,
  in Euclidean signature, is \cite{BST,Sezgin}
\be
S_{M2}=\tau_{M2}\int_{\Sigma}d^3s \Bigl [ {\sqrt{detg_{ij}}}+
if^*{\bf{ C}^{(3)}}
\Bigr ]- i\tau_{M2}\int_{\partial \Sigma}d^2\sigma\phi^*B^{(2)}.
\label{M2}
\ee
Here  $\tau_{M2}=\frac{1}{(2\pi)^2}M_{11}^3$ is the M2-brane tension.
Also,  $B^{(2)}$ is the super 2-form living on ${\cal M}$ while ${\bf
C}^{(3)}$
is the super 3-form  living on the target superspace $M.$
The   pullback in eq.(\ref{M2}) under the 
embedding $f: s^i \rightarrow
Z^{\underline M}$  is 
$$f^*{{\bf C}^{(3)}}=\frac{1}{3!}\partial_iZ^{\underline M}
\partial_jZ^{\underline N}
\partial_k Z^{\underline P}
 {\bf C}^{(3)}_{{{\underline M}}{{\underline N}}{\underline P}}ds^i\wdg
ds^j
 \wdg ds^k, \quad $$
while the pullback under the embedding $\phi: \sigma^r \rightarrow z^M$
is
$$\phi^*B^{(2)}=\half
\partial_r z^M \partial_s z^N B^{(2)}_{MN}d\sigma^r \wdg d\sigma^s,
\quad $$

We specialize the  action (\ref{M2}) to the case of 
 an M2-brane stretched between $y=0$ and $y=x$ in the background described 
in section 2. The membrane is a product $\Sigma \times [0,x]$ so it 
 is convenient to define coordinates on the membrane 
$s^i=(t,\sigma, \bar{\sigma})$ where $t$ is a coordinate on the 
interval and $\sigma$ is a holomorphic coordinate along the 
curve $\Sigma$. The embedding coordinates of $\Sigma$ into $(11\vert 32)$
superspace 
\be
Z^{\underline M}(s)=\Bigl (X_{3,11}^{\hat M}(s), \Theta_{3,11}(s) \Bigr ),
\label{embd}
\ee
have the following structure. First, the interval coordinate
\be
\displaystyle
X_{3,11}^{11}(s)=\pi \r \Bigl (t+\frac{i}{2} \Theta_{3,11}\Gamma^{11}
{\overline \Theta_{3,11}} \Bigr ) 
\label{x11}
\ee
has an important correction quadratic in fermions, while the coordinates
\be
X_{3,11}^{m}(s)=X^{m}(\sigma), \quad
X_{3,11}^{\bm}(s)=X^{\bm}(\bar{\sigma}), 
\label{e311}
\ee
describe the holomorphic embedding. The coordinate $X_{3,11}^{\mu}(s)$ is 
unconstrained. The fermions $\Theta_{3,11}(s)$ satisfy the physical gauge
condition
\be
\Gamma^{(2)}\Theta_{3,11}(s)=-\Theta_{3,11}(s).
\label{phys}
\ee
We have omitted the coordinates describing 
fluctuations of the membrane within  $\cX$ since we 
will restrict our consideration to an isolated curve 
$\Sigma$ and hence these degrees of freedom will be massive. 

The origin of the correction in (\ref{x11}) is continuity of 
embedding coordinates in superspace. That is, the 
embedding of the membrane into 11-dimensional 
superspace $(3|0)\rightarrow (11|32)$ must agree, on the 
boundary, 
with the embedding of the 5-brane into superspace 
$(6|16)\rightarrow (11|32)$ since the membrane ends on 
the 5-brane {\it in superspace}. 
We now derive this condition in more detail. 

 We choose    bosonic coordinates on the M5-brane
as $y^M=(y^{\mu},y,\by)$ where $y^\mu$ are real and 
$y$ is complex, and consider the static embedding
of the boundary of the M2-brane into the M5-brane
 \be
\phi: y=\sigma, \by =\bar{\sigma}
\label{boundary}
\ee
 The superembedding $(6|16)\rightarrow (11|32)$
is described by superfields 
$$Z^{{\underline M}}=\bigl ( X^{\hat M}(z^M), \Theta(z^M) \bigr ) $$
Small fluctuations around
static gauge are described by embeddings
\be
X^{\hat M}=\Bigl (y^m,X^{m'}(y,\vth)\Bigr ),
 \quad \Theta=\Bigl (\vth , \psi(y,\vth) \Bigr )
\label{flct}
\ee
where $\vth$ is a 
fermionic coordinate in the $(6\vert 16)$ superspace.
The superfields $X^{m'}(y,\vth)$ and $\psi(y,\vth)$ 
have as their $\vth=0$ component bosonic fluctuations 
transverse to the worldvolume of the M5-brane $X^{m'}(y^M)$
 and physical
fermions on the M5-brane $\psi(y^M)$ respectively. $m'$ here denotes
bosonic indices of coordinates transverse to the M5-brane.
 
As was discussed, for example,  in \cite{Sorokin}, the
basic superembedding condition (\ref{supemb})
imposes a relation on the superfields $X^{m'}(y,\vth)$ and $\psi(y,\vth),$ 
such that 
\be
X^{m'}(y^M,\vth)=X^{m'}(y^M)+i\vth \Gamma^{m'}\psi(y)+...
\label{relation}
\ee
In particular, the superfield $X^{11}$ up to linear order in $\vth$ is
\be
\displaystyle
 \quad
X_{6,11}^{11}=\pi \r \Bigl (x+i \vth \Gamma^{11} {\overline \psi}+...
\Bigr )
\label{e611}
\ee
Recall that we introduced the factor $\pi \r$ to make $x$ dimensionless.

In the geometry of    $\cX$, the spinors  $\vth$ and $\psi$ can be
decomposed
as
\be
{ \vth}=\Bigl \{\vth^{\dot \a}\otimes\epsilon_1,
{\vth}^{  \a}\otimes\epsilon_2 \Bigr \}.
\label{vth}
\ee  
\be
{ \psi}=\Bigl \{\psi^{ \a}\otimes\epsilon_1,
{\psi}^{\dot  \a}\otimes\epsilon_2 \Bigr \},
\label{psi}
\ee  
Out of  the 16-component spinors we only keep those 
components given by the covariantly constant 
spinor along $\cX$.  There are other physical 
degrees of freedom in the spinor $\psi$, but since 
we are considering a {\it rigid} curve  in 
$\cX$ only the above components lead to massless 
degrees of freedom.  

In Euclidean space 
 equation (\ref{e611}) becomes
\be
\displaystyle
X_{6,11}^{11}=\pi \r \Bigl (x-\frac{i}{R}
\Bigl( \vth^{\dot \a} \psi_{\dot \a}+ 
\vth^{ \a} \psi_{ \a}\bigr )\Bigr )
\label{e611new}
\ee
where now chiral and anti-chiral spinors are independent
from each other.
(To give the meaning to the fermionic bilinears in Euclidean
space, we first define them in Minkowskii space,
where
\be
\displaystyle
\vth_{\dot \a}
=\bigl(\vth^{\a} \bigr)^*, \quad
\psi_{\dot \a}=
\bigl(\psi^{\a} \bigr)^*,\quad \gamma^0=\Biggl (\begin{array}{cc}
 0 &1 \\
1 &0 \\
\end{array} 
\Biggr )
\label{Mink}
\ee
where the   spinors indices are lowered 
via $\vth_{\dot \a}:=\varepsilon_{{\dot \a}{\dot \b}}\vth^{\dot \b}.$
Then we continue to Euclidean signature by dropping the reality
conditions in eq.(\ref{Mink}).)

{}From eq.(\ref{x11}) and eq.(\ref{e611new}) we see that
$X_{3,11}^{11}$ and $X_{6,11}^{11}$ match each other
at the boundary of the M2-brane, i.e. at $t=x,$
if the following boundary conditions are imposed on the physical
fermions $\Theta_{3,11}$

\be
\Theta_{3,11}^{{\dot \a }{\dot Y}}|_{t=x}=
\vth^{\dot \a}\otimes\e_1^{\dot Y}, \quad
\Theta_{3,11}^{{\dot \a} Y}|_{t=x}=
\psi^{\dot \a}\otimes\e_2^{ Y},
\label{brycond}
\ee
where $\dot Y$($Y$) are chiral( anti-chiral) spinors indices on $\cX.$

One can identify  zero modes living
on the boundary of the M2-brane
$\vth^{\dot \a}\otimes\e_1^{\dot Y}$
with the supersymmetry broken by the   M2-brane.  
This is compatible
with our considerations in section 3. Indeed, exactly these   
components of background supersymmetry are  broken for the instanton 
described by the embedding  eq.(\ref{x11}). 

The bosonic part of the M2-action is
\be
\displaystyle
 S_{M2}=Z, \quad   Z= R\b_ia^i x -i\hcA,
\quad \hcA=\cA(\b_ia^i)-x(\b_i\chi^i).      
\label{z}
\ee
where, as in section two, $\beta_i [\Sigma^i_2]$ is the homology 
class of the boundary curve. 

 Now, by evaluating the embedding matrices for an M5-brane 
up to  linear order in $\vth$ and solving 
    the equation in the $(6|16)$  superspace ( see \cite{Sezgin})
 \be
 dB=H+f^{*}{\bf C},
\label{Sezgin}
\ee
we obtain the expression for $B^{(2)}_{y \by}$ up to  linear order
 in  $\vth$   
\be 
B^{(2)}_{y \by}=A^{(2)}_{y \by}+i{\partial_y}X^n
{{\partial}}_{\by}X^{\bm}\Bigl(
-\vth{\Gamma}_{n\bm}{\overline \psi}+
C_{n\bm {\hat P}}\vth{\Gamma}^{{\hat P}}{\overline \psi}\Bigr).
\label{super}
\ee

In solving equation (\ref{Sezgin}) we have used constraints 
on the superform $H.$ Specifically, we have used the condition 
that the only non-vanishing components of the superform $H$, in the basis
$e^A$,
are the components $H_{abc}$ with all three bosonic indices.
This was derived in \cite{Sezgin} by requiring
$\kappa$-supersymmetry of the action of an M2-brane ending on an
M5-brane.

In (\ref{super}) we have  dropped      
terms containing derivatives of the fluctuating fields 
such as  $\partial_{\mu}\cA$, since these terms in the 
vertex operator will 
not contribute to the fermion two-point function 
we wish to compute. 

Now, from eq.(\ref{M2}) and eq.(\ref{super}),
 and using the properties (\ref{norma})
 of covariantly constant spinors on  $\cX,$
 we evaluate the interaction between zero-modes of
  fermions living on an M2-brane boundary
 $\vth_{\dot \a}$ and  fermions $\psi_{\dot \a}$
to be
\be
\displaystyle
V_{\psi}=i(\b_ia^i)\vth^{\dot \a }\psi_{\dot \a}
\label{vert1}
\ee

Note that the contribution to the interaction
 from the second boundary term in eq.(\ref{super}) was cancelled
by the term from the bulk, after integrating by parts,
 due to the presence of the piece in the embedding (\ref{x11})
which was quadratic in fermions.

The last step in deriving the ``vertex operator'' for
the chiral fermion  superpartner of $Z$, denoted $\chi_{\dot \a},$
is to  relate $\psi_{\dot \a}$ and $\chi_{\dot \a}.$
To achieve this  we consider a supersymmetric  
variation of $Z$ with supersymmetry parameter
$(\epsilon^{\dot \a}\otimes \e_1,\epsilon^{ \a}\otimes \e_2). $ 
 The 
result is
\be
\displaystyle
 \delta Z=(R\b_ia^i)\delta x +x\delta (R\b_ia^i)-i\delta \hcA 
\label{varZ}
\ee
where
\be
\displaystyle
\delta x =i{ \e}\Gamma^{(11)}{\overline \psi}=
-\frac{i}{R}\Bigl (\epsilon^{\dot \a    }\psi_{\dot \a}+
\epsilon^{ \a }\psi_{ \a}\Bigr )
\label{varx}
\ee
\be
\displaystyle
\delta \hcA =(\b_ia^i)\Bigl (\epsilon^{\dot \a }\psi_{\dot \a}-
\epsilon^{ \a }\psi_{ \a}\Bigr )
\label{varA}
\ee
 Equation (\ref{varA})
is a direct consequence
of eq.(\ref{super}) and the definitions (\ref{defA},\ref{z}).

Denoting by $\lam^i_{\dot \a}$ the superpartners of the bulk scalars
$T^i,$ we get the desired relation
\be
\displaystyle
\chi_{\dot \a}=2(\b_ia^i)\psi_{\dot \a }+x(\lam^i_{\dot \a } \b_i)
\label{relation}
\ee
and hence the ``vertex operator'' for $\chi_{\dot \a}$ is
\be
\displaystyle
V_{\chi}=\frac{i}{2}\vth^{\dot \a }\chi_{\dot \a}.
\label{coupl}
\ee  
 
\section{  The case of one M5-brane.  }

In this section we will discuss the scalar potential 
for  the case of 
one M5-brane located at position $y=x$ on the 
M-theory interval. The general formula 
has been quoted above in (\ref{U0}). In order to 
evaluate this expression we need both $K$ and $W$. 
We will describe first $W$ and then $K$, and then 
put them together.

\subsection{Superpotential }

In the present setting the superpotential
$ W $ can be written as a sum of 5 pieces
\be
W=W_{\rm pert}+W_2+W_3 +W_4 +W_5.
\label{W}
\ee
which have the following origins: 

\begin{itemize}
\item{ $W_{\rm pert}$ is the Yukawa superpotential for the charged chiral
superfields,
given in (\cite{GSW,Ovrut2})
\be
\displaystyle
W_{\rm pert}=\frac{(4\pi) \sqrt{2}}{3}\lam_{\hI_1 \hI_2 \hI_3}C^{\hI_1}
C^{\hI_2} C^{\hI_3}, 
\label{W1}
\ee
The Yukawa couplings are given by
\be
\lam_{\hI_1 \hI_2 \hI_3}=\int\Omega\wdg u^{x_1}_{I_1}\wdg 
u^{x_2}_{I_2} \wdg u^{x_3}_{I_3} f_{x_1x_2x_3}^{(\cR_1 \cR_2 \cR_3)}
\label{Yuk}
\ee
  and  depend on the  complex structure and bundle moduli,
but are independent of $T^i$ and $S.$ $f_{x_1x_2x_3}^{(\cR_1 \cR_2
\cR_3)}$
projects onto the singlet in $\cR_1\times \cR_2\times \cR_3$ 
( if it exists), and $\Omega$ is a choice of nowhere zero 
holomorphic $(3,0)$ form on $\cX$. 
 
}

\item{
$W_2$ is the sum of two pieces coming from an M2-brane
stretched between the M5-brane and the boundary 9-brane at $y=0$ or 
$y=1$ respectively.
We have shown in the previous section that 
\be
\displaystyle
W_2 = h \Bigl \{ exp\Bigl( - Z \Bigr)+ 
exp\Bigl( - (\b_iT^i - Z) \Bigr) \Bigr \}
\label{W2}, 
\ee
}
(Note that the relative sign can be changed by a shift in 
the imaginary part of $T$ or $Z$.) 
 
\item{
$W_3$  is the contribution to the superpotential 
 due to gaugino condensation studied in  \cite{Hor,Ovrut2,Dine}.
It is given by

\be
W_3 \sim exp\Bigl( -\frac{3}{2b_0 } S \Bigr ),
\label{W3}
\ee
where $b_0$ is a beta-function coefficient for the gauge
group on the second 9-brane. 
 
We are working  in  a region of   moduli space
constrained by (\ref{eff}). It follows that 
\be
3V\gg 2b_0  Ra^i\b_ix, \quad  3V\gg 2b_0 Ra^i\b_i(1-x)
\label{ass}
\ee
and hence the contribution of $W_3$ to the potential
is much smaller than the contribution of $W_2$, and will 
henceforth be neglected.

}
\item{ $W_4$ is the contribution  from    M2-brane  instantons 
stretched between the two
``M9-branes.''
The contribution   from a single membrane wrapping a 
holomorphic curve $\Sigma\subset \cX$  has the form

\be
\displaystyle
W_4 \sim exp\bigl( - \b_iT^i \bigr), 
\label{W4}. 
\ee
In the 
case of the  standard embedding ( with  no 5-branes)
the sum over all such curves in a fixed homology class  
vanishes.   This happens because   $W_4$ is just the 
world-sheet instanton contribution to the   superpotential 
 in the effective theory near a 
 (2,2) vacuum of 
the weakly coupled heterotic string.
 Such  superpotentials for moduli are  known to be zero
(\cite{Cand,Strom,Wit}).
$W_4$ is also zero for the special cases of the ``non-standard embedding''
arising in     weakly coupled  heterotic (0,2) vacua which 
are related to linear sigma models. For example,
$W_4=0$ for the quintic in $P^4.$ 
Nevertheless, it is expected that these corrections will 
be nonzero for   generic  $(0,2)$ compactifications \cite{Den1,Den2}. }

\item{$W_5$ is the superpotential coming from an M5-brane wrapping the
whole
$\cX.$ 
\be
W_5\sim \exp\Bigl(-\tau_{M5} Sv\Bigr) =  \exp\Bigl( -\frac{1}{4} S
 \Bigr )
\label{W5}. 
\ee
}
$W_5$ is of the same order as $W_3$ and again we 
can neglect it relative to the effects of open 
membranes.

\end{itemize}

\subsection{Kahler potential for bulk moduli and charged scalars}

To evaluate the scalar potential in (\ref{U0}) one also needs  the Kahler
potential
\footnote{Here we are assuming that the moduli space is a product space, 
as is valid in our approximation. } 
\be
K= K_S+ K_T+ K_m+K_{cplx}+K_5 + K_{\rm bundle},
\label{K}
\ee

The first four pieces in this expression have already been obtained in
previous papers. We will derive a formula for $K_5$ below. It would 
be interesting to learn more about $K_{\rm bundle}$, but we will not 
do so in this paper. In this section we review the results for the 
first four terms, obtained in \cite{Ovrut1,Ovrut2,Li,Dud,Nil}.

The first two terms in (\ref{K}) are: 

\be
\displaystyle
 K_S=-ln(S+\bS),\quad 
K_T=-ln\Bigl(\frac{1}{6}d_{ijk} (T^i+\bT^i)(T^j+\bT^j)(T^k +\bT^k)\Bigr )
\label{K0}
\ee
To leading order in an expansion in $C^I$ the charged matter has a
K\"ahler 
potential of the form

\be
K_m=Z_{\hI \hJ} C^{\hI}{\overline C}^{ \hJ} + \cdots 
\label{K_m}
\ee
Here $Z_{\hI \hJ}$ is constructed from  the  metric for bundle moduli
$G_{B \hI \hJ}$ as follows. First of all, $G_B$ is defined by 
\be
G_{B \hI \hJ}  =\frac{1}{vV}\int_{\cX}\sqrt{g }{g}^{m\bm}
u_{I m x} u^x_{J \bm} 
\label{bundle}
\ee
and depends on the Kahler moduli $a^i,$ as well as on the 
complex structure
and bundle moduli.
Next we define 
$K_{B \hI \hJ}:=e^{\frac{K_T}{3}} G_{B \hI \hJ }$. Note that the 
dependence on the Kahler moduli is only through the combination 
$ T^i+\bT^i$.
 Then we can define
\be
\displaystyle
Z_{\hI \hJ}=G_{B \hI \hJ}
 - \frac{e^{-\frac{K_T}{3}}2\xi_i}{S+\bS}
{\tilde \Gamma}^i_{B\hI \hJ}
\label{Z}
\ee
where  $\xi_i$ was defined in (\ref{b_j}), and
\be
\displaystyle
 \Gamma^i_{B\hI \hJ}=K_T^{ij}
\frac{\partial   K_{B \hI \hJ }}{\partial T^j},
\quad K_{Tij}=\frac{\partial^2K_T}{\partial T^i\partial {\bT}^j}
\label{bundle0}
\ee

$${\tilde \Gamma}^i_{B\hI \hJ}= \Gamma^i_{B\hI \hJ}-(T^i+\bT^i)
 K_{B \hI \hJ}
-\frac{2}{3}(T^i+\bT^i)(T^k+\bT^k)K_{Tkj} \Gamma^j_{B\hI \hJ},$$
$K_T^{ij}$ 
denote the inverse of the matrix $K_{Tij}.$

In formulae (\ref{Z}) and (\ref{K_m})  the following restrictions on the
scalar
fields are assumed 
\be
 Z_{\hI \hJ}C^{\hI}{\overline C}^{\hJ} \ll 1
\label{yes}
\ee
\be
\frac{2 \xi_i}{S+\bS}
{\tilde \Gamma}^i_{B\hI \hJ} \ll K_{B\hI \hJ}.
\label{no}
\ee

 The Kahler function for the complex structure moduli is
 \be
K_{cplx}=-ln\Bigl ({\overline{\Pi^a}}{\cal{G}}_a\Bigr ), 
\quad {\cal{G}}_a= \partial_{\Pi^a}{\cal{G}}, \quad a=1, \ldots, h_{21}+1
\label{Kcom}
\ee
 and can be expressed in terms of the periods over the A-cycles $\Pi^a$
and the prepotential  ${\cal{G}}$, with complex structure moduli 
expressed as $\pi^{\a}=\frac{\Pi^{\a}}{\Pi^0},\quad a=(0,\a).$
 
\subsection{Kahler potential for the M5-brane moduli }
 
The last piece in (\ref{K}) is $K_5,$ the Kahler potential giving the 
kinetic terms for
the 5-brane scalars $x$ and $\cA.$
 As we were finishing our project 
we found that (\cite{Der}) obtained $K_5$ in the special case
of $h^{(1,1)}=1.$
Since we got our result independently and in a  different way,
we will explain the derivation below.

To find  $K_5$ we start from  the bosonic part of the
 Pasti-Sorokin-Tonin action for the M5-brane \cite{PST}

\be
S_{M5}=\tau_{M5}\int_{W_6}d^6y\Biggl
(-{\sqrt{-det(\gamma_{MN}+iH_{MN})}}
-\frac{1}{4}{\sqrt{-\gamma}}v_LH^{*LMN}H_{MNP}v^P \Biggr )
\label{PST}
\ee
$$+\tau_{M5}\int_{W_6}\Biggl ({\hat C_6}+
\half dA_2\wdg {\hat C_3} \Biggr ),$$
Here the tension of the M5-brane is $ \displaystyle
\tau_{M5}=\frac{1}{(2\pi)^5}M_{11}^6 .$
The other terms in the action are defined by
$$\displaystyle \gamma_{MN}=\frac{\partial X^{\hat M}}{\partial y^M }
\frac{\partial X^{\hat N}}{\partial y^N }g^{(11)}_{{\hat M}{\hat N}},
\quad H^{MN}=H^{*MNP}v_P, \quad H^{*MNP}=-\frac{1}{3!{\sqrt{-\gamma}}}
\epsilon^{MNPLKQ}H_{LKQ}$$
\be
H_{MNP}=3\partial_{[M}A^{(2)}_{NP]}-({\hat C_3})_{MNP}, \quad
{{\hat C}}_{MNP}=\frac{\partial X^{\hat M}}{\partial y^M }
\frac{\partial X^{\hat N}}{\partial y^N }
\frac{\partial X^{\hat P}}{\partial y^P }C_{{\hat M}{\hat N}{\hat P}}
\label{fivedefs}
\ee
$$\displaystyle
 {{\hat C}}_{M_1 \ldots M_6}=\frac{\partial X^{\hat M_1}}{\partial y^{M_1}
}\ldots\frac{\partial X^{\hat M_6}}{\partial y^{M_6} }C_{{\hat M_1}\ldots
{\hat M_6}} \quad
v_N=\frac{\partial_N \Phi
}{\sqrt{\partial_K \Phi
\partial^K\Phi}},\quad v_Nv^N=1.$$
where 
$\Phi$ is the PST scalar and $\hat C_6$ is the magnetic dual of $\hat C_3$.

We wish to do Kaluza-Klein reduction of the above action along the 
holomorphic curve $\Sigma$. We split the coordinates in the
bulk as $X^{{\hat M}}=(\xi^{\mu}, X^a,X^{\ba},X^{11})$ 
where $a,\ba =1,\ldots,3$ are indices for the complex coordinates, 
and $\xi^\mu$ are coordinates along the noncompact $R^4$. 
We choose $\mu$ to run over $\mu=0,1,2,5$. 
\footnote{We have changed notation from section two for this 
computation.} 
The coordinates along the worldvolume $W_6$ of the 5-brane 
are taken to be $y^M$ which we split as 
4 real coordinates  $y^\mu$, $\mu=0,1,2,5$ and one 
complex coordinate $y$ along the holomorphic curve $\Sigma$. 
A natural gauge choice for the PST scalar is  \cite{Sch}
\be
v_M=\delta_M^5,\quad A_{5M}=0 
\label{gauge1}
\ee
While the gauge choice (\ref{gauge1}) breaks  
six dimensional covariance, after the KK reduction 
we will obtain a   covariant 4-dimensional action. 

The  massless
fluctuations of the M5 brane are described by fields 
$$X^{11}=\pi \r x(\xi^{\mu}), \quad A_{m n}(\xi^{\mu}),
\quad  \cA(\xi^{\mu}),\quad \mu=(m,5),\, m=0,1,2$$ 
where $\cA(\xi^{\mu}) $ was defined in equation (\ref{defA}).
Keeping only terms of quadratic order in derivatives 
we obtain the following 
4-dimensional action\footnote{ We are grateful to Steuard Jensen
for pointing out a mistake in the original version
of the equation below.}
\be
\displaystyle
S_{M5}=-  v^{\frac{1}{3}}\tau_{M5}\int_{W_4}\Biggl \{
\half(\pi \r)^2 \frac{ e(a^i\b_i)R}{V}(\partial_{\mu}x)(\partial^{\mu}x)
-\half \frac{ (a^i\b_i)RV}{eg^{55}}(H^{y\by})^2
\label{W4}
\ee
$$\displaystyle 
+\half (\pi \r)^2\frac{e}{(a^i\b_i)RV}\Bigl
[\partial_m\hcA+x\partial_m(\b_i\chi^i)\Bigr ]
g_{(3)}^{mn}\Bigl [\partial_n\hcA+x\partial_n(\b_i\chi^i)\Bigr ]$$
$$\displaystyle
+(\pi \r)\frac{g^{5m}}{g^{55}}\Bigl
[\partial_m\hcA+x\partial_m(\b_i\chi^i)\Bigr ]H^{y\by}
+(\pi \r)\Bigl [\partial_5\hcA+x\partial_5(\b_i\chi^i)\Bigr ]H^{y\by}
+(\pi \r)^2\frac{ex}{V^2}\partial^{\mu}\sigma \partial_{\mu}\Bigl [
\cA(\b_i a^i)\Bigr ] \Biggr \}$$
where  $e={\sqrt{-detg_{\mu \nu}}},$ while
$g^{5m}$ and $g^{55}$ are components of the inverse of the 4-dimensional 
 metric  $g^{\mu \nu}.$ One should take care that  
$g_{(3)}^{mn}$ is the inverse of the 3-dimensional metric so that
$$\displaystyle 
g^{mn}=g_{(3)}^{mn}+\frac{g^{5m} g^{5n}}{g^{55}}.$$
Moreover, in (\ref{W4}) we have
$$H^{y\by}=\half \epsilon^{mnp}\Bigl
[\partial_mA_{np}-\partial_mX^{11}C_{11np}
\Bigr ],$$
(where we have used (\ref{fivedefs})) and finally we 
 have also introduced
$$ \hcA=\cA(\b_ia^i)-x(\b_i\chi^i),\quad
\partial^m\sigma=g^{m\nu}\partial_{\nu}\sigma .$$

One can see from (\ref{W4}) that integrating out $H^{y \bar y}$ gives
$$\displaystyle
H^{y\by}=(\pi \r)\frac{e}{(\b_ia^i)RV}g^{5\mu}\Bigl [
\partial_{\mu}\hcA+x\partial_{\mu}(\b_i\chi^i)\Bigr ]$$
Plugging the expression for $H^{y \by}$ back into (\ref{W4})
restores  4-dimensional covariance
and results in the action
\be
\displaystyle
\kappa_4^2 S_{M5}=-\int_{W_4}e\Biggl \{
\half \frac{ (a^i\b_i)R}{V}(\partial_{\mu}x)(\partial^{\mu}x)
+\half\frac{1}{(a^i\b_i)RV}\Bigl
[\partial_{\mu}\hcA+x\partial_{\mu}(\b_i\chi^i)\Bigr ]^2
\label{W41}
\ee
$$+\frac{x}{V^2}\partial^{\mu}\sigma \partial_{\mu}\Bigl [
\cA(\b_i a^i)\Bigr ]\Biggr \}$$

One can now extract the Kahler potential for the 5-brane moduli. 
The terms in the action (\ref{W41}) uniquely determine   $K_5$ 
to be
\be
K_5= \frac{(Z+\bZ)^2}{(S+\bS)(\b_iT^i+\b_i\bT^i)},
\label{K5}
\ee
A check of the supergravity kinetic terms associated with   $K_5$ shows
that 
but there are extra terms coming from (\ref{K5}) 
and given by: 

\be
-  \int_{W_4}e\Biggl \{
-x(a^i\b_i)RV^{-2}(\partial_{\mu}x)(\partial^{\mu}V)
-\half x^2 V^{-2}(\partial_{\mu}\Bigl (R(a^i\b_i)\Bigr )(\partial^{\mu}V)
\label{extra}
\ee
$$-\frac{x^2}{2V^2}\partial_{\mu}(\b_i \chi^i)(\partial^{\mu}\sigma)
- \frac{x(\b_i\chi^i)}{ V^2}\partial_{\mu}x\partial^{\mu}\sigma 
+\half x^2(a^i\b_i)RV^{-3}\Bigl (\partial_{\mu}V \partial^{\mu}V +
\partial_{\mu}\sigma \partial^{\mu}\sigma \Bigr ). \Biggr \}
$$
These terms are exactly cancelled by the terms coming from
$K_S=-ln(S+\bS)$ after including an $x-$dependent correction to
 the definition of
the chiral field $S$ 

\be
 S=V+i\sigma+ T^i\b_i x^2
\label{SS}
\ee

Note, that the above correction $ T^i\b_i x^2$ is of 
order $\ce$
with respect to $V.$ 
There are no $x-$dependent corrections to the other
fields at this order.

The proper interpretation of these facts is that the complex structure 
on field space is corrected at the nonlinear level by (\ref{SS}) 
and that the K\"ahler potential $K_S + K_5$ should be written as 
\be 
\widehat{K_S} = - \log \biggl[ S+ \bS - \frac{(Z+\bZ)^2}{
(\b_iT^i+\b_i\bT^i)} \biggr]
\ee
It would be interesting to learn if this expression is valid at higher 
order in the expansion in $Z$.

\subsection{ Potential in the case $h^{(1,1)}=1.$ }

In the previous sections we have given formulae 
valid   for a generic $\cX.$ 
We will   now specialize to 
the case of $h^{(1,1)}=1$ in order to simplify the analysis
of the potential. As we have stressed above, in this case  
 there is no net contribution to the non-perturbative
potential from M2-branes stretched between the two boundary 9-branes,
i.e. $W_4=0.$

When $h^{(1,1)}=1$ the dependence  on the
Kahler parameter $a$  
of the  metric (\ref{bundle}) for 
the bundle moduli can be easily extracted by 
a scaling argument.  We can choose a basis  
 $u_{I m}^x $ that does  not depend
on $a$. Then   the inverse of Calabi-Yau metric scales 
like 
$$\displaystyle g_{CY}^{m \bm}=\frac{1}{a}\omega_{(1)}^{m \bm}$$
where   $\omega_{(1),n \bm}$ is, say, an integral 
generator of  $H^{(1,1)}(\cX)\cap H^2(\cX,Z).$
Under these conditions 
the  Kahler metric for charged scalars (\ref{bundle0})
 simplifies considerably
 and is given by
\be
\displaystyle
Z_{\hI \hJ}=\Bigl ( \frac{3}{T+\bT} + \frac{2 \xi}{S+\bS}\Bigr )
H_{\hI \hJ}, \quad \xi=\b^{(0)}+\b (1-x)^2,
\label{quintic}
\ee
where $H_{\hI \hJ}$ depends only on complex structure  
and bundle moduli.  
In the case
at hand $ H_2(\cX,Z) $ is of rank $1$ and generated by a 
rational curve $\Sigma$.  We  take $\b=1,$ which
corresponds to wrapping a 5-brane only once around $\Sigma.$

The perturbative potential for charged scalars was obtained in
\cite{Ovrut2}. 
Using the formulae from the appendix we have calculated
 the non-perturbative potential including explicit leading
dependence on Kahler moduli and    charged scalars.  
As mentioned in the introduction we write the 
full potential in the form 
\be
\displaystyle
{(\kappa_4)}^{4}U=
\Bigl (U_{0}+U_{1} + U_{2} \Bigr )
\label{U}
   \ee
where, 
as mentioned in the introduction, we organize terms 
by the order in the nonperturbative superpotential. 
$U_{0}$ begins with the perturbative 
potential. $U_{1}$ results from 
mixing between the perturbative and nonperturbative 
contributions to $W$, while $U_2$ is the term
of second order in the nonperturbative 
superpotential.    The formula for the potential 
contains a prefactor $e^K$. We have used the explicit results 
for $K_S$ and $K_T$, and 
we have dropped $K_m$ and $K_5$ since they contribute subleading 
effects to the order we are working.

We now give the leading expressions for the 
three terms in (\ref{U}) in more detail.  
The leading contributions to $U_{0}$  are given by 
\be
\displaystyle
U_{0}=\frac{4\pi^2}{3 \td}
 \frac{  e^{K_{\rm cplx} + K_{\rm bundle}} }{V  J^2}
|\lam C C|^2
+ U_D
\label{Upert}
\ee
In the above formulae
  $ V={\tilde d}a^3, \quad J:=Ra,$
where $d=6\td$ is the intersection number on $\cX.$
The expression
\be
\displaystyle
|\lam C C|^2=
C^{\hI_1}C^{\hI_2}\lam_{\hI_1 \hI_2 \hI_3}H^{\hI_3 \hJ_3}
{\overline{\lam}}_{\hJ_1 \hJ_2 \hJ_3}{\bC}^{\hJ_1}{\bC}^{\hJ_2}, \quad
\label{Upert1}
\ee
 comes from derivatives of $W$
with respect to $C^{\hat I}$.

The $D$-term is given by 
\be
U_D= \frac{18 \pi^2   }{ V J^2  }\sum_{(a)} \Bigl( {\bC}^{\hI} H_{\hI \hJ}
T^{(a)} C^{\hJ}
\Bigr )^2 
\label{Dterm}
\ee
where  $T^{(a)}$ are generators of
the  unbroken four dimensional gauge group $H$, and we are assuming 
there are no induced FI parameters.

To leading 
order $U_1$ is
given by: 
\be
\displaystyle
U_{1} =-\frac{  e^{K_{\rm cplx} +
 K_{\rm bundle}}(1-x)}
{2 \td V J^2 }
\Biggl \{
 e^{-Jx} \Re\Bigl(W_{\rm pert} \bar h e^{i \alpha_1} \Bigr) 
- 
 e^{-J(1-x)} \Re\Bigl(W_{\rm pert} \bar h e^{i \alpha_2} \Bigr) 
\Biggr\}
\label{U2}
\ee
 where we define the axion fields 
\be
\displaystyle
\a_1=ImZ, \quad \a_2=Im(T-Z), 
\label{not3}
\ee
There will be corrections from terms higher order in 
the   expansion in $\frac{\vert C \vert ^2}{J},\ce,\cer$. There are also 
corrections from multiply-wrapped membranes. 
 
The leading contribution to $U_2$ is a two-instanton term
\be
\displaystyle
U_2=\frac{e^{K_{\rm cplx} + K_{\rm bundle}} }{8 \td J^2}
 |h|^2 \Bigl \{ e^{-2Jx}+e^{-2J(1-x)} -2e^{-J}\cos\bigl (\a_1-\a_2 \bigr )
\label{U1}
\ee
$$+ \frac{2J}{3V}\bigl (1-2x\bigr )e^{-2J(1-x)}+
\frac{4Jx}{3V}e^{-J}\cos\bigl (\a_1-\a_2 \bigr )\Bigr \}
  +\cdots $$
The leading piece comes from $K^{Z\bZ}\vert \partial_Z W\vert^2$. 
Note that in the second line of (\ref{U1}) we have kept terms 
which are formally higher order in our expression since they 
multiply $J/V\sim \ce$. We have kept these because, near $x=1/2,
\quad cos\bigl (\a_1-\a_2 \bigr )=1$ the 
leading piece vanishes. At these points the order $J/V$ corrections  
which come from $K^{Z\bZ}$ and the prefactor $e^K$    multiply zero and we
 can legitimately say that  the leading term near $x=1/2,
\quad \cos\bigl (\a_1-\a_2 \bigr )=1 $
 is given by 
the last term in the second line of (\ref{U1}).
 We will need these subleading 
terms in section 5.5 below. 
 Of course, there are 
many other corrections of  relative order $ \cO\bigl ((\ce)^p (\cer)^q
,\frac{\vert C \vert^2}{J}
\bigr )$, where $p\geq 1$ and $q>0$.     
 
The region of validity of our result for the potential,  (\ref{U}), 
is constrained by several considerations.

\begin{itemize}
\item 
We must assume that all sizes are much bigger than the 11D Plank length.

\be
 \pi \r Rx \gg l_{11}, \quad \pi \r R(1-x) \gg l_{11}, 
 \quad a^{\half}v^{\frac{1}{6}} \gg l_{11}
 \label{c1}
 \ee
and
from these conditions it follows, in particular, that 
$Jx \gg 1, \quad J(1-x) \gg 1.$

\item Since we are working to quadratic order in the K\"ahler potential 
in a series expansion in $C$ we must have
\be
|C|^2:= C^{\hI}H_{\hI \hJ}{\bC}^{\hJ}\ll J
\label{c3}
\ee

\item   
The effective expansion parameters should be small,
or, equivalently, 
\be
V \gg  J ,\quad J^2 \gg  V
\label{c3}
\ee

\item 
We must be able to  drop $\ce$ corrections  to each of the 9 terms
 in the potential. We count all the terms wich have different
sructure,i.e.
5 from $U_2,$ 2 from $U_{1}$ and 2 from $U_{0}.$

$$U=\sum_{a=1}^{9}u_a(1+ f_a \ce),$$

iff

$$ \ce |\sum_{a=1}^{9}f_a u_a | \ll |\sum_{a=1}^{9} u_a | $$
Given our ignorance regarding   $f_a$  we  will assume    that they
are $O(1)$ and 
impose  the stronger condition 
\be
 \ce \sum_{a=1}^{9} |u_a | \ll |\sum_{a=1}^{9} u_a | 
 \label{c7}
 \ee

\item 
Finally, as mentioned in the introduction, we should stress that 
we are assuming that we are working at a generic smooth point in 
the complex structure and bundle moduli space.

\end{itemize}

\subsection{5-brane dynamics} 

We can get some heuristic idea about the 5-brane 
dynamics  by considering the 
theory on a finite volume of 3-dimensional space 
and keeping only the spatially homogeneous modes of the 
scalar fields. Even in this drastic approximation 
the resulting system is a very complicated 
dynamical system described by a particle mechanics 
Lagrangian with the (very) schematic form  
\be
{\rm vol(space)}
\int dt \Biggl\{ \bigl( \frac{\dot V}{V}\bigr)^2 + 
\bigl( \frac{\dot J}{J}\bigr)^2 +  \frac{(\dot C)^2}{J} +
\frac{(\dot J x + \dot x J)^2}{V J} - U\Biggr\}
\ee
where the potential energy is 
\be 
U=
\frac{1}{V J^2} \Biggl(\alpha C^4 - \beta(1-x) \vert C\vert^3 \vert
 e^{-J x} \mp 
 e^{-J(1-x)} \vert +   \qquad\qquad\qquad 
\label{schempot}
\ee
$$+\gamma V \Bigl [ \bigl(e^{-Jx} \mp e^{-J(1-x)}\bigr )^2+
\frac{2J}{3V}\bigl (1-2x\bigr )e^{-2J(1-x)}
\pm\frac{4Jx}{3V}e^{-J} \Bigr ]\Biggr)$$ 

We have  only kept the real 
parts of the fields. The philosophy behind this is 
that by a   ``Born-Oppenheimer'' type approximation 
we expect that the axions will relax much more rapidly than 
the real parts into the most attractive 
channel. The choice of $\pm$ depends on what
term is dominating, $U_1$ or $U_2.$

The positive constants $\alpha,\beta$ and $\gamma$
are functions of the complex structure and bundle 
moduli, but these are being held fixed in this discussion. 

While the dynamical system we must study is rather 
complicated we can get some heuristic idea of 
what to expect in three distinct regions within 
the region of validity of our potential. 

\begin{itemize} 

\item

In one region the charged scalar fields are zero, 
 while $J$ and $V$ are large. The leading contributions 
to the potential are positive and decrease with 
increasing $J,V$.   
In this region the 5-brane 
leads to a {\it repulsive} interaction between 
the M9 walls. Setting $C=0$ and choosing ``- `` sign
to set axions in the most attractive channel
in eq.(\ref{schempot}) we get: 
\be
U \sim \gamma \frac{1}{J^2}\Biggl \{
 \bigl (e^{-Jx} - e^{-J(1-x)}\bigr )^2+
\frac{2J}{3V}\bigl (1-2x\bigr )e^{-2J(1-x)}+
\frac{4Jx}{3V}e^{-J}\Biggr\}
\label{smooth}
\ee
Note that  $U$ has a minimum in $x$ 
at  $x=\half.$
Expanding around
the minimum   $x=\half +y,$ the 
resulting potential is
\be
\displaystyle
U =
\gamma \Bigl \{\frac{2}{3VJ}e^{-J}+4 y^2e^{-J}\Bigr \}.
\label{potval}
\ee

Now we can see the need for keeping  
 the last two terms in the expression
for $U_2$ in eq.(\ref{U1}).  Although we   are neglecting the
 $\ce$-corrections to the Kahler
potential,
  such corrections  multiply the leading  three terms
in eq.(\ref{U1})
which sum up to zero at $x=\half.$ On the other hand, 
the terms we have written, and which follow from 
the leading pieces in $K$    become the leading 
terms 
at the stationary points $x=\half.$ 
Consequently,   $J$ flows towards 
infinity and $x$ moves towards the middle of 
the interval. 
We must assume 
$ V^{1/2} \ll J \ll V $
to stay within the  region of validity.

\item
If the charged scalar fields are  important in such a
way that 
 \be
\vert U_1 \vert \gg U_2,
\label{regionII}
\ee
 then the leading $x$-dependent 
term in the potential is {\it attractive}:
 $$U =\a C^4 - \beta(1-x) \vert C\vert^3 \Bigl(
 e^{-J x} + 
 e^{-J(1-x)} \Bigr)$$
Note that in this case we have to choose the ``+'' sign 
in eq.(\ref{schempot})  
if the axions are in the most attractive channel. 
Thus, if $x<1/2$ the 5-brane moves towards 
the wall at $y=0,$
and if $x>1/2$ the 5-brane moves towards the wall at $y=1.$
 Note, that  
 in each of these two subregions   $U_0 \gg \vert U_1\vert$
as a direct consequence of (\ref{regionII}). Since 
$U_0$ is the dominant term $J$ and $V$ will evolve to large
values. Since  
 the $C$ field is 
simultaneously evolving a more 
careful analysis of the dynamical system 
would be highly desirable. But we  will not do that here. 
 
\end{itemize}
 More generally, one can show that 
  the potential 
(\ref{schempot}) is    non-negative at a generic point
in the bundle 
and complex strucure moduli space, within 
in our  region of validity (\ref{c1}-\ref{c7}), and thus 
predicts decompactification
of both the  Calabi-Yau and the orbifold interval. 
The argument, which is straightforward but long can 
be found in Appendix B. 

Note that (\ref{schempot}) is the leading potential
 only under our assumption
that we work at a generic smooth point in  
bundle and complex structure moduli space. 
It would  be interesting
to incorporate  singularities in 
complex structure and bundle moduli space in the discussion.
There are potentially many new terms in the 
potential that must be reconsidered.   
It is possible that using the known results 
on complex structure and bundle moduli space one can address 
this problem.

\subsection{Conflicting instabilities}

One interesting consequence of the discussion in the 
previous section is that there is a strong coupling 
dual of the Dine-Seiberg problem where the M-theory interval 
(and the Calabi-Yau) tends to decompactify. In the 
case of heterotic M-theory with the standard embedding 
(i.e. no 5-branes) this has already been discussed 
by Banks and Dine \cite{Dine2}, who noted 
that  one can use holomorphy to extrapolate 
the weak coupling superpotential based on gluino condensation. 
In the presence of an M5-brane (in the case $ h^{(1,1)}=1$)
the above formulae show that in 
the region  specified by (\ref{c1}-\ref{c7})
there is a similar effect 
due to open membrane instantons.

It is of some interest to compare the above result with 
what we expect for the weakly coupled heterotic string, 
since our considerations are only valid at large 
heterotic string coupling. Indeed, the heterotic 
coupling is related to the length of the M-theory 
interval by
$$\displaystyle R \r \sim l_{11}(g_s)^{\frac{2}{3}}$$
 and we require
 $\pi \r R \gg l_{11}.$
In the   regime of weak coupling and   large $V$,  the 
potential  has been discussed in  \cite{DineSeib}.
It was shown there that the effective potential
is positive and behaves as
\be 
\displaystyle
U_{eff}\sim e^{-\frac{V}{g_s^2}}.
\label{small}
\ee
This favors an evolution to weak coupling $g_s \to 0$ and 
large volume $V \to \infty$. 
One might worry that the calculations of (\cite{DineSeib})
 were performed in the case of the standard embedding, 
and in backgrounds with other $E_8 \times E_8$ 
 gauge instantons  one must
 take into account the contribution of world-sheet
 instantons as well \cite{Den1,Den2}. Nevertheless, as
 we have repeatedly mentioned, 
these effects often sum to zero \cite{Silv,Kach} so 
once again  and we can use (\ref{small}) in the region of small $R$
  and large $V.$           
 
In view of the above, we can combine   our result (\ref{U}) with
(\ref{small}),
 to  learn that the "true" potential
   goes to zero through positive values 
in both   limits  $R\rightarrow 0$
 and $R\rightarrow \infty.$
 This indicates that there must  be a stationary point somewhere
 in the intermediate region, i.e.
 at some finite value of $R.$
The nature of the stationary points  that lie in the middle 
of moduli space is unknown, and is, of course, an interesting 
and outstanding 
question.

\section{Multiple covers and chirality changing transitions} 

It is of considerable interest to determine the nature of the 
low energy theory in the limit that the M5 moves into the 
the boundary 10-manifold. In this section we will make some 
comments on this limit. We will need to make some guesses 
and the results of this section are not as rigorous as in 
the previous sections. For definiteness we will consider the 
limit $x\to 0$. 

One good reason for studying
 the limit $x\to 0 $ is  that there 
are strong indications that in such limits there can be very interesting 
chirality-changing phase transitions in the low energy theory. 
This was discovered, in the present context, by Kachru and Silverstein
\cite{Kachru2}. The theory of these transitions has been 
considerably extended to many new  examples in \cite{smallinst}. 

The main new ingredient that is needed to discuss the $x\to 0$ 
behavior is a multiple-cover formula for the open membrane 
instantons. The fact that there must be nontrivial effects 
from multiply-wrapped M2 branes (at least for those
stretching between two 5-branes) can be seen by considering 
the  holomorphy of the full superpotential
$W$ as a function of  $Z_i - Z_j$ \cite{HM}. The instanton effects 
must be suppressed by a factor proportional to the 
volume of the stretched membrane and therefore 
must behave like $\exp[\mp (Z_i - Z_j) ] $ for 
$\pm Re(Z_i-Z_j) > 0$. This is only consistent with 
holomorphy if there is an infinite series with at 
best a finite radius of convergence.

Multiply-wrapped worldsheet instanton corrections to $d=4, N=2$ 
prepotentials are known to have a universal form 
$f(n,\Sigma)$ for an $n$-times wrapped curve $\Sigma$
where $f$  only depends on the topology of $\Sigma$ 
\cite{CandGreen, AspMorr, Morrison, Marino, faber, Vafa4}. 
Since worldsheet instantons are special 
cases of M2-brane instantons we will make the working 
hypothesis that there is similarly a multiple cover factor 
$f(n,\Sigma)$ for M2-brane instanton corrections 
to the superpotential $W$. Some  evidence for 
this can be found in \cite{ Vafa3, Kachru1, Vafa2}. 
Unfortunately, the topologies studied 
in the above papers do not contain our case of $P^1 \times [0,1]$. 
Therefore we will take 
\be
\displaystyle
\Delta W=h \sum_{n=1}^{\infty}f(n) e^{-nZ}+ e^{i \delta}h \sum_{n=1}^{\infty}f(n) e^{-n(T-Z)}
\label{sum}
\ee
and make the rather 
weak assumption that  
the asymptotic behavior of $f(n)$ for large 
$n$ is $f(n) \sim n^m e^{J x_0 n}$ for some constants $m$ and $x_0$.
For simplicity we set $x_0=0$ although there could 
in principle be a shift in the location of the small instanton 
transition.

The constants $m$ and $\delta$ above are unknown, 
but we can make some comments on them. 
First, the  relative phase $e^{i \delta}$  
was not important in the   1-instanton sector,
where we can change the relative phase by
shifting the axion $ImT.$ It does become a nontrivial
issue in the multi-instanton sectors. Nevertheless, for our
analysis of the dynamics in the subregion $ReZ\ll 1$
the second piece in (\ref{sum}) is negligible,
so the issue need not concern us
here.

Next, let us consider the  power $m$ in the asymptotic 
behavior of $f(n)$. 
If we wish the chiral fermion mass term in standard supergravity 
 to be  a single-valued function 
of $Z$ in a region surrounding $Z=0$, then 
$m$ cannot be a negative integer such that $\vert m \vert > 2.$ 
Single-valuedness 
implies that the  monodromy of 
$\partial_Z \partial_Z W$ around $Z=0$ 
should be diagonalizable thus excluding  
a singularity of the type $Z^n \log Z, n \geq 2$ in $W.$   
\footnote{There are known examples of logarithmic 
superpotentials for n=0,1 that make good physical sense. This is 
usually related to some kind of pair creation phenomena. 
We thank K. Hori for very useful discussions on this issue. }

Let us now re-consider the region of validity of our expression.
The infinite series for  $\Delta W$
can be obtained reliably in the region  $ReZ \gg 1 $ 
(where we can use 11-dimensional supergravity) and then 
analytically continued to the region
 where $ReZ=Jx \ll 1.$ 
To ensure that corrections to the Kahler function
 are small one must still require that
$$V \gg 1,\quad V^{\half} \ll J \ll V,\quad \vert C\vert^2 \ll J $$ 
Since these conditions do not imply     
$Jx \gg 1$ or $J(1-x) \gg 1$ we can study the physics of 
the 5-brane approaching the boundary.

For definiteness and simplicity let us now assume that 
$f(n) = n^m$ for some constant $m$. Then we have
\be
\partial_Z \Delta W=-h 
 Li_{-(m+1)}\bigl (e^{-Z}\bigr )
\label{dw}
\ee
where  $Li$ is a polylogarithm function
$$Li_{-(m+1)}(t)=\sum_{n=1}^{\infty}n^{m+1} t^n
$$

In this case the leading 
order contribution to $U_1$ is
given by: 
\be
\displaystyle
U_{1} =-\frac{  e^{K_{\rm cplx} +
 K_{\rm bundle}}(1-x)}
{2 \td V J^2 }
 \Re\Biggl [ \bW_{\rm pert}  h
  Li_{-(m+1)}\bigl (e^{-Z}\bigr )
 \Biggr ]
\label{nU1}
\ee
while the leading contribution to $U_2$ is 
\be
\displaystyle
U_2=\frac{e^{K_{\rm cplx} + K_{\rm bundle}} }{8 \td J^2}
 |h|^2 {\Biggl \vert}Li_{-(m+1)}\bigl (e^{-Z}\bigr)
{\Biggr \vert}^2 .
\label{nU2}
\ee

In all the cases below we will assume 
 that the axion phases are in the
 maximally attractive channel.

One interesting limit is $Z\to 0$. Here we can 
  use the behaviour of the polylogarithm 
$$Li_{-(m+1)}(e^{-Z}) \sim Z^{-(m+2)}, \quad Z \rightarrow 0 $$
(for $m=-2$ we replace $Z^0$ by $\log Z$)  
to write out (schematically ) the leading potential at $Jx \ll 1$
\be 
U=
\frac{1}{V J^2} \Biggl(\alpha C^4 - \beta(1-x) \frac{\vert C\vert^3}
{(Jx)^{2+m}}
+\gamma \frac{V}{(Jx)^{4+2m}} \Biggr ), \quad m >-2
\label{nschempot}
\ee

\be 
U=
\frac{1}{V J^2} \Biggl(\alpha C^4 +\beta  \vert C\vert^3(1-x)
log(Jx)
 +\gamma V \bigl (log Jx \bigr )^2  \Biggr ), \quad m=-2 
\label{npot4}
\ee
where $\a, \b, \g $ are positive functions
of complex structure and bundle moduli as above and 
we have  set   $ImZ=0.$

Therefore, for small enough $Jx$ (holding the other moduli 
fixed) 
the leading term in the potential (\ref{nschempot} ),(\ref{npot4})
is  
\be
 \frac{\g }{J^2 (Jx)^{4+2m} }, \quad  m >-2, \qquad\qquad
 \frac{\g (logJx)^2}{J^2  }, \quad m=-2
\label{leading}
\ee
and there is a repulsive force on the 5-brane. Indeed there is  an infinite
energy barrier forbidding the 5 brane from hitting the wall.

One should not conclude from the above that there will be 
no chirality changing transition, since the axionic degree 
of freedom in $Z$ can change the qualitative features of 
the potential drastically. Unfortunately, in order to study 
this question in detail just knowing 
the asymptotic behavior will not suffice 
and one needs a precise version of
the formula for $f(n)$ in order to work out 
the analytic continuation from $Re(Z)\gg 1$ to 
$\vert e^{-Z} \vert =1$. For definiteness 
we will consider   
$f(n) = n^m$ where  $m \geq -2 $ is integral. 

Let us consider first the case $m\geq 0$. 
We take $Z=Jx +i\pi,$ where $Jx \ll 1$
and expand $e^{-Z}=-1+Jx -\half(Jx)^2.$
We use the Taylor expansion for the polylogarithm
$$ Li_{-(m+1)}(-1+t)= Li_{-(m+1)}(-1)-
 Li_{-(m+2)}(-1)t+\half\Bigl ( Li_{-(m+3)}(-1)- Li_{-(m+2)}(-1)\Bigr) t^2 $$
and the following useful relations
$$ Li_{-m}(-1)=(2^{1+m}-1)\zeta(-m), \quad
\zeta(-2k)=0, \quad \zeta(1-2k)= - \frac{B_{2k}}{2k}, \quad k=1,2,\ldots
$$
where  $\zeta$ is the Riemann $\zeta$-function and $B_k$ are 
Bernoulli numbers taken in the convention:
$$\frac{y}{e^y-1}=1-\half y +\frac{B_2}{2!}y^2+\frac{B_4}{4!}y^4 +...$$
Substituting $t=Jx -\half(Jx)^2$ in the above
Taylor expansion and keeping only terms up to $(Jx)^2$
we have
\be
Li_{-(m+1)}(-1+t)=(-1)^{k+1}\Biggl (\nu_1^k-\nu_2^k(Jx)^2 \Biggr ),  \quad
m=2k, \quad k=0,\ldots
\label{even}
\ee

\be
Li_{-(m+1)}(-1+t)=
(-1)^{k+1}2\nu_2^kJx,
\quad m=2k+1, \quad k=0, \ldots
\label{odd}
\ee
where we define positive numbers $ \nu_1^k, \nu_2^k$ 
$$ \nu_1^k=\bigl ( 2^{2k+2}-1 \bigr )\frac{\vert B_{2k+2}\vert }{2(1+k)},
\quad \nu_2^k=\bigl ( 2^{2k+4}-1 \bigr )\frac{\vert B_{2k+4}\vert }{4(2+k)}. $$ 

We now analyze the potential separately for  the cases of even and odd $m$. 
 For even $m=2k$ we have the following leading potential
at $Jx \ll 1$ 
\be 
U=
\frac{1}{V J^2} \Biggl(\alpha C^4 - \beta(1-x) \vert C\vert^3 \Bigl (
\nu_1^k- \nu_2^k (Jx)^2 \Bigr )
+\gamma V\Bigl (
\bigl (\nu_1^k \bigr )^2-  2\nu_1^k\nu_2^k (Jx)^2 \Bigr ) \Biggr ) 
\label{npot1}
\ee
If $C\not=0$ then for sufficiently small $x$
 the potential is attractive. If $C=0$ the potential is 
repulsive.

Now, for odd $m=2k+1$   the leading potential  is
\be 
U=
\frac{1}{V J^2} \Biggl(\alpha C^4 - 2\beta\nu_2^k(1-x)Jx \vert C\vert^3
 +\gamma V
\bigl (2\nu_2^k\bigr )^2 (Jx)^2  \Biggr ) 
\label{npot2}
\ee
Now the situation is opposite to the previous case. 
For $C\not=0$ there is a repulsive force and if 
$C=0$ an attractive force. 

Finally we analyze what happens for the cases  
 $m=-1$ and $m=-2$, assuming $ImZ=\pi $ and $Jx \ll 1.$
If $m=-1$ the leading potential is
\be 
U=
\frac{1}{V J^2} \Biggl(\alpha C^4 - \half
\beta  \vert C\vert^3(1-x)\Bigl (1-\half Jx\Bigr )
 +\frac{\gamma V}{ 4}\Bigl (1- Jx\Bigr )  \Biggr ) 
\label{npot3}
\ee
The force on the 5-brane is attractive
only if one allows for a large vev for $ \vert C\vert^3$
$$ \b \vert C\vert^3 > \g V .$$
This is in principle possible
since we  only assume that  $\vert C\vert^3 \ll V^{\frac{3}{2}}$
and for large $V$ both inequalities can be satisfied.
If  $m=-2$ the  leading potential   is
\be 
U=
\frac{1}{V J^2} \Biggl(\alpha C^4 - 
\beta  \vert C\vert^3(1-x)\Bigl (log2-\half Jx\Bigr )
 +\gamma V log2\Bigl (log2- Jx\Bigr )  \Biggr ) 
\label{npot3}
\ee
and the force is attractive only if
$$ \b \vert C\vert^3 > 2 (log2)\g V .$$
In both cases $m=-1$ and $m=-2$,
attraction is only possible for large vev of
charged scalars.

The general conclusion based on the analysis of various
cases  is that the 
  physics of what happens when the 5-brane 
approaches the wall depends strongly on the detailed form 
of the multiple-cover formula. 

Finally, let us comment on the relevance of this compututation 
to the examples studied by Ovrut, Pantev, and 
Park in  \cite{smallinst}. One might 
at first conclude that in these examples   the superpotential must 
vanish  since the five-brane wraps a high 
genus curve.  However, the curve wrapped by the 5-brane is not 
irreducible and not isolated. It can very well happen 
that in the long-distance expansion of the M5 and M2 
Lagrangians there are terms with many fermions (typically
multiplying factors involving curvature tensors) which 
can lift the many fermion zeromodes associated with the 
nonisolated high genus curve. Thus, the question of whether 
or not a superpotential is generated is a complicated and 
difficult one, involving a discussion of the measure 
on the moduli space of the curve and the integral over that 
moduli space. Considerations based on the global form of 
the moduli space for these curves based on the results 
of \cite{DonagiOW} do not appear to exclude the generation 
of such superpotentials.

\section{The case of N M5-branes. }

We will now briefly consider the potential in the case that there are 
N M5-branes at positions $x_1< x_2<\ldots < x_N$.
We will assume for simplicity that all the 5-branes are wrapped
over the same rational curve $\Sigma$,
 so that open M2-instantons can be stretched
between any pair of  5-branes. Moreover, to 
simplify the analysis we assume that 
the 5-branes are more or less evenly
separated.
Finally, we restrict our consideration only to the leading
non-perturbative potential, so we do not take into account 
2-instanton contributions to $W$  and we need only 
consider M2-branes
between neighboring 5-branes. Similarly, we only
keep the contribution of 5-9 instantons coming from
M2-instantons stretching between the boundary and the 
nearest M5-brane. 
Under these conditions we will have 
$$ R(x_n-x_{n-1})\pi \r \gg l_{11}, \qquad \forall n=1,\ldots, N+1.$$

Neglecting $\ce$ corrections due to the distortion
of the background, the K\"ahler function for the collection of 5-branes
 will be just a sum of Kahler functions for each 5-brane.

The potential is again given by formula (\ref{U}), with the 
same conditions on the region of validity.  The 2-instanton 
terms in the potential $U_{2}$, which dominate at $C=0$, are: 
\be
U_{2} =\frac{e^{K_{\rm cplx} + K_{\rm bundle}}}{8 \td J^2} |h|^2 
\sum_{n=1}^{N}
 \Bigl \{e^{-2J(x_{n+1}-x_n)}+e^{-2J(x_n-x_{n-1})}
-2e^{-J(x_{n+1}-x_{n-1})}
cos\bigl ({\tilde \a}_n \bigr ) \Bigr \}+\cdots
\label{UUUU}
\ee
where we denote
\be
\displaystyle
{\tilde \a}_n= \Biggl \{ a\bigl
(2\cA_n-\cA_{n+1}-\cA_{n-1}\bigr )
 + \chi \bigl (x_{n+1}+x_{n-1}-2x_n \bigr ) \Biggr \},n=1, \ldots,N
\label{alpha0}
\ee
and $x_0=0,x_{N+1}=1,\cA_0=\cA_{N+1}=0.$
If instead we assume that
\be
x_1\gg x_n-x_{n-1}, \quad (1-x_N)\gg x_n-x_{n-1},\quad
 \forall 1<n\le N
\label{toda}
\ee
and choose  a special subregion where $\cos{\tilde \a}_n=0,\quad \forall
n,$
then the potential has the form of  a non-periodic Toda-chain potential. 
(The kinetic energies are the standard ones, in our approximation.)
As is well known, Toda theory has an exact solution,
where all particles move away from each other \cite{perelomov}. 
In   heterotic M-theory
 this signals an instability in the time evolution of the 
positions of M5-branes along the orbifold interval: they 
 tend to run away from 
each other.  At the same time $Ra$ evolves to infinity. 
In short, the system explodes. 

Using again a   ``Born-Oppenheimer'' type approximation 
we expect that the axions will relax much more rapidly than 
the real parts into the most attractive 
channel  $cos{\tilde \a}_n=1,\quad \forall n.$
This implies that the evolution with a Toda-like potential is 
unstable because of the axions.  

When the charged vevs are nonzero we should consider instead 
the term  $U_{1}$ in  the potential arising from cross 
terms between perturbative and nonperturbative pieces. This is given 
by
\be
\displaystyle
U_{1}=-\frac{   e^{K_{\rm cplx} +
 K_{\rm bundle}}\vert h\vert \z}
{2\td J^2 V}
\Biggl \{
e^{-Jx_1}(1-x_1) cos({\tilde\g}_1)
-e^{-J(1-x_N)}(1- x_N)
cos({\tilde\g}_N) 
\label{UUU}
\ee
$$-\sum_{n=1}^{N-1}(x_{n+1}-x_n) 
e^{-J(x_{n+1}-x_n)} cos({\tilde\phi}_n)  \Biggr \}$$
 where $  W_{\rm pert}=\z e^{i \phi_1}, h=\vert h \vert e^{i \phi_h}$ 
are decompositions into modulus and phase and 
\be
\displaystyle
{\tilde\g}_1=ImZ_1+\phi_1-\phi_h, \quad 
{\tilde\g}_N=Im(T-Z_N)+\phi_1-\phi_h, 
\label{lu}
\ee

$${\tilde\phi}_n=Im(Z_{n+1}-Z_n)+\phi_1-\phi_h,
\quad n=1, \ldots, N-1$$

\section{Possible future directions and applications}

A central question in heterotic M-theory is the 
existence of isolated  minima of the potential 
for moduli. While most of our results predict 
runaway or unstable behavior (as expected) 
we have seen some encouraging hints. 
We have argued that the potential must have 
nontrivial stationary  points in moduli space. 
We have also seen that a good place to look for 
interesting behavior of the potential is at singular loci in 
complex structure and bundle moduli space. For example, 
if one 
allows some of the coefficients $\alpha,\beta, \gamma$ 
in sections 5.5 and section 6 to vanish it is easy 
to imagine scenarios where the potential predicts 
compactification, rather than decompactification.

There are many technical issues 
raised by the above computations
which should be solved and which moreover  can be 
solved with presently available technology. 

One circle of questions  includes 
finding the appropriate generalization 
of the multiple-cover formula for worldsheet instantons. 
A related set   of questions concerns effects associated with membranes 
wrapping higher genus curves and nonisolated curves in $\cX$. 
As we have seen in section six, results on these questions 
would have very interesting physical applications.

A second circle of questions concerns the possibility of 
obtaining   a more concrete understanding 
of the dependence of the membrane determinants as functions of 
the complex structure moduli. It might   be possible to 
find classes of compactifications in which one can give fairly 
explicit formulae for the dependence on gauge bundle moduli, 
although this might prove to be challenging.

Beyond the extensions mentioned above, 
which we believe are within reach, there 
loom far more difficult questions. One of the most challenging 
issues is to give a proper definition of Horava-Witten theory in a 
regime outside the validity of the expansion in $(\kappa_{11})^{2/3}$. 
  Another difficult, and pressing, problem is that of 
finding ways to make definite and quantitative statements about the 
K\"ahler potential of the effective supergravity theory in a 
wider range of validity. 

Nevertheless, even given the limitations of our computations, 
the results do have some interesting ranges of validity. It might 
be quite interesting to study more thoroughly the dynamics, 
both classical and quantum mechanical of the moduli in the problem. 
In this paper we have limited ourselves to some very heuristic and 
naive pictures of the dynamics. 
It might also be interesting to see if there are any 
distinctive features of the ``modular cosmology'' resulting 
from the above potential for moduli \cite{modcosmo}.

\vskip 0.5in
\centerline{\bf Acknowledgements} 

We would like to thank Jeff Harvey and Marcos Mari\~no for 
collaboration at some stages of  this research.  
We would also like to thank  B. Acharya, 
T. Banks, D.-E. Diaconescu,
 M. Douglas,  S. Kachru, O.Ganor, S. Gukov, 
M. Mari\~no,
A.Mikhailov, N.Nekrasov, and M.Rangamani 
  for   discussions
and useful correspondence.  The work
of GM  is supported by DOE grant DE-FG02-96ER40949.
The work of GP is supported by the Rutgers DOE grant DE-FG02-96ER40949
and by the Yale DOE grant DE-FG02-92ER-40704.

\section{Appendix A}

For the convenience of the reader we list here   the leading expressions
for Kahler potential
in the case $h^{(1,1)}=1$, together with formulae for the Kahler metric
and 
inverse metric.  The Kahler potential is:
\be
 K= K_S+ K_T+ K_m+K_{cplx}+K_5 + K_{\rm bundle},
\label{appK}
\ee
$$ K_S=-ln(S+\bS),\quad 
K_T=-ln\Bigl(\td \bigl ( T+\bT \bigr )^3 \Bigr )
$$
$$K_5= \frac{(Z+\bZ)^2}{(S+\bS)(\b_iT^i+\b_i\bT^i)}$$

$$K_m=\Bigl (\frac{3}{T+\bT}+\frac{2\xi}{S+\bS}\Bigr )H_{\hI \hJ}
 C^{\hI}{\overline C}^{ \hJ}$$ 

We now give the components of the Kahler
metric on the space of scalars which have been used in section 5.4.
We keep only leading terms  
 in each of the component, neglecting corrections of the relative
order $O\bigl(\ce,\cer,
\frac{\vert C \vert^2}{Ra} \bigr).$

$$\displaystyle K_{S \bS}=\frac{ 1 }{4V^2}, \quad
K_{S \bT}=\frac{ x^2}{4V^2}, \quad
K_{S  \hJ}=-\frac{ \xi H_{\hI  \hJ }
C^{\hI}}{2V^2}, \quad
K_{S {\overline \a}}=-\frac{ \xi \partial_{{\overline \a} }
H_{\hI {\overline \hJ} } C^{\hI} {\bC}^{{\overline \hJ}} }{2V^2},  
$$
$$\displaystyle
K_{S \bZ}=-\frac{ x }{2V^2}, \quad
K_{T \bT}=\frac{3}{(2Ra)^2}, \quad
K_{ \bT I}=-\frac{3 H_{\hI \hJ}{\bC}^{\hJ}}{(2Ra)^2}
$$
$$
K_{ \bT \a}=-\frac{3 C^{\hI}\partial_{\a}H_{\hI \hJ}{\bC}^{\hJ}}{(2Ra)^2}
\quad K_{Z \bT}=-\frac{  x }{2V R a}, \quad
K_{\hI {\hJ}}=\frac{3}{2Ra}H_{\hI {\hJ}}, \quad
K_{\hI {\overline \a} }=\frac{3 
\partial_{{\overline \a}}H_{\hI {\hJ} }{\bC}^{{\hJ}}}{2Ra}$$

$$\displaystyle
K_{Z {\hI} }=- \frac{  (1-x)
H_{{\hI} \hJ }C^{\hJ}}{V Ra}, \quad
K_{\bZ \a }=- \frac{ (1-x)C^{\hI}
\partial_{\a}H_{\hI {\hJ} }{\bC}^{{\hJ}} }
{V Ra}$$
$$ \displaystyle 
K_{\a {\overline \b} }=K^{(cplx)}_{\a {\overline \b} },\quad
K_{Z \bZ}=\frac{1}{2VRa}$$

Now, we solve the matrix equation
$$ K K^{-1}=1 +O\bigl (\ce,\cer, \frac{\vert C \vert^2}{Ra} \bigr )$$
The inverse metric  solving this equation is

$$\displaystyle
K^{S \bS}=4V^2, \quad
K^{T \bT}= \frac{(2Ra)^2}{3},
\quad
K^{ \bT \hJ}=C^{\hJ} \frac{2Ra}{3},
$$

$$\displaystyle
K^{\hI {\hJ }}=\frac{2Ra}{3}H^{\hI \hJ }, \quad
K^{{\overline \a} \b}= K_{cplx}^{{\overline \a}\b}, \quad
K^{{\hJ} \b}=-\partial_{{\overline \delta} }H_{\hI
{\hK}}
  {\bC}^{{\hK} } H^{\hI  {\hJ} }K^{{\overline
\delta} \b} $$
$$ \displaystyle 
K^{ Z \bZ}=2RaV, \quad
K^{ Z \bS}=4RaVx,\quad K^{ Z \bT}=\frac{(2Ra)^2}{3}x, \quad 
K^{\hJ \bZ }=\frac{(2Ra)}{3} C^{\hJ}\bigl (2-x \bigr ) $$
where the components not listed above are zero in our 
approximation.

\section{Appendix B} 

In section 5.5 we asserted that, within the region of validity of 
our computations, the potential is always positive. Here we give 
the detailed proof of that claim.

The only potentially negative term in the potential is 
$U_1$. We will show that it cannot be larger
in magnitude than both of $U_0$ and $U_2$ in our 
region of validity. 

First, 
imposing $$\vert U_1 \vert \geq  U_2$$ 
means
\be
\beta (1-x) |C|^3\Bigl ( e^{-Jx}+e^{-J(1-x)} \Bigr ) \geq
\gamma V \Bigl(e^{-Jx} + e^{-J(1-x)}\Bigr )^2
\label{prf1}
\ee
 It follows immediately that
$$ \displaystyle |C| \geq 
\Bigl (\frac{\g V}{\b}\bigl [ e^{-Jx}+e^{-J(1-x)} \bigr ]\Bigr
)^{\frac{1}{3}}.$$

Now, at a generic point in bundle and complex
moduli space, we have
 $$\displaystyle \a C^4 \geq
\a |C|^3\Bigl (\frac{\g V}{\b}\bigl [ e^{-Jx}+e^{-J(1-x)} \bigr ]\Bigr
)^{\frac{1}{3}}\gg \b |C|^3(1-x)\Bigl ( e^{-Jx}+e^{-J(1-x)} \Bigr )$$ 
and we see that  $U_0 \gg \vert U_1\vert.$

Let us now  assume  $$\vert U_1 \vert \geq U_0.$$ 
{}From this it follows that
$$ \b (1-x)\Bigl ( e^{-Jx}+ e^{-J(1-x)} \Bigr ) \geq \a \vert C \vert $$
and hence
\be
 \displaystyle |U_1| \leq \frac{1}{J^2 V} \frac{\b^4}{\a^3}
(1-x)^4 \Bigl ( e^{-Jx}+ e^{-J(1-x)} \Bigr )^4
\label{prf3}
\ee

 Let us consider, first,the region far enough from $x=1/2.$
Then, for $x<\half,$ we have
$$\displaystyle |U_1| \leq \frac{1}{J^2 V} \frac{\b^4}{\a^3}
 (1-x)^4e^{-4Jx}$$
and
$$\displaystyle U_2 \sim \frac{1}{J^2} e^{-2Jx}$$
As a consequence, $U_2 \gg \vert U_1\vert.$

In the region close to $x=\half$  
we have instead, for sign ``+'' in eq.(\ref{prf3})
$$\displaystyle |U_1| \leq \frac{1}{J^2 V} \frac{\b^4}{\a^3}e^{-2J}$$
and
$$\displaystyle U_2 \sim \frac{1}{J^2} e^{-J}$$
and  it follows immediately that
 $$U_2 \gg \vert U_1\vert.$$
For sign ``-'' in eq.(\ref{prf3}) the last statement  is obvious.



\begin{thebibliography}{99}
\bibitem{HorWit}{P. Ho\v rava, E. Witten, 
``Heterotic and Type I String Dynamics from Eleven Dimensions''
      Nucl.Phys. B460 (1996) 506-524, hep-th/9510209}
\bibitem{HorWit1}{P. Ho\v rava, E. Witten,
''Eleven-Dimensional Supergravity on a Manifold with Boundary'',
 Nucl.Phys. B475 (1996) 94-114,hep-th/9603142}
\bibitem{Witten}{E.Witten,
`` Strong Coupling Expansion Of Calabi-Yau Compactification'',
Nucl.Phys. B471 (1996) 135-158,hep-th/9602070}
\bibitem{Dine}{T.Banks and M. Dine, 
''Phenomenology of Strongly Coupled Heterotic String Theory'',
Nucl.Phys.B479(1996)173.hep-th/9609046}
\bibitem{Dine2}{T.Banks and M. Dine, 
''Couplings and Scales in Strongly Coupled Heterotic String Theory'',
(hep-th/9605136)}
\bibitem{Witten2}{E.Witten, ``Non-Perturbative Superpotentials 
In String Theory'',  Nucl.Phys. B474 (1996) 343-360,hep-th/9604030}
\bibitem{BBS}{K.Becker, M.Becker, A.Strominger, Five-branes, Membranes
 and Non-perturbative String Theory.
Nucl.Phys.B456:130-152,1995,hep-th/9507158 }
\bibitem{HM}{J. Harvey, G.Moore, Superpotentials and 
Membrane Instanons, hep-th/9907026}
\bibitem{plefka}{B. de Wit, K.Peeters, J. Plefka,
       `` Open and Closed Supermembranes with Winding,''
        Nucl.Phys.Proc.Suppl. 68 (1998) 206-215}
\bibitem{wittenDinst}{ E. Witten,World-Sheet Corrections Via D-Instantons,
hep-th/9907041}
\bibitem{shamit}{S. Kachru, S. Katz, A. Lawrence, J. McGreevy,Open string
instantons and superpotentials Phys.Rev. D62 (2000) 026001; 
hep-th/9912151}
\bibitem{acharya}{ B.S. Acharya, M theory, Joyce Orbifolds and Super
Yang-Mills,hep-th/9812205 
}
\bibitem{Den1}{M.Dine, N.Seiberg, X.G.Wen, E.Witten,
``Nonperturbative effects on the string worldsheet'',
Nucl.Phys. B278(1986)769-789}
\bibitem{Den2}{M.Dine, N.Seiberg, X.G.Wen, E.Witten,
``Nonperturbative effects on the string worldsheet (II)'',
Nucl.Phys. B289(1987)319-363}
\bibitem{Hor}{ P. Ho\v rava ,
'' Gluino Condensation in Strongly Coupled Heterotic String Theory'',
Phys.Rev. D54 (1996) 7561-7569,hep-th/9608019}
\bibitem{Ovrut1}{ A. Lukas, B. A. Ovrut, D. Waldram, 
``Non-standard embedding and five-branes in heterotic M-Theory'',
 Phys.Rev. D59 (1999) 106005, hep-th/9808101  }
\bibitem{Ovrut2}{ A. Lukas, B. A. Ovrut, D. Waldram,
"Five-branes and Supersymmetry Breaking in M-theory",
 JHEP 9904 (1999) 009;hep-th/9901017 }
\bibitem{Ovrut3}{ A. Lukas, B. A. Ovrut, D. Waldram,K.Stelle,
'' Heterotic M-Theory in Five Dimensions,hep-th/9806051'',
 Nucl.Phys. B552 (1999) 246-290     }
\bibitem{Ovrut4}{R. Donagi, A. Lukas, B. A. Ovrut, D. Waldram
  `` Non-Perturbative Vacua and Particle Physics in M-Theory'',
        JHEP 9905 (1999) 018, hep-th/9811168 }
\bibitem{Ovrut5}{R. Donagi, A. Lukas, B. A. Ovrut, D. Waldram,
Holomorphic Vector Bundles and Non-Perturbative Vacua in M-Theory''
       JHEP 9906 (1999) 034,hep-th/9901009}
\bibitem{Silv}{E.Silverstein, E.Witten,
``Criteria for Conformal Invariance of (0,2) Models'',
 Nucl.Phys. B444 (1995) 161-190,hep-th/9503212 }
\bibitem{Kach}{J.Distler, S.Kachru,''Singlet Couplings and (0,2) Models'',
       Nucl. Phys. B430 (1994) 13, hep-th/9406090}
\bibitem{candelas3}
{ P. Berglund, P. Candelas, X. de la Ossa, E. Derrick, J. Distler, T.
Hubsch,
 On the Instanton Contributions to the Masses and Couplings of $E_6$
Singlets,Nucl. Phys. B454 (1995) 127;hep-th/9505164
}
\bibitem{Cand}{P. Candelas, X. de la Ossa, Moduli Space of Calabi-Yau
Manifolds. Nucl.Phys.B355:455-481,1991
}
\bibitem{Strom}{A.Strominger,'' Special Geometry'', 
Comm.Math.Phys.1333(1990)163.}
\bibitem{Wit}{E.Witten, ``Phases of N=2 Theories in two Dimensions'',
Nucl.Phys. B403(1993)159.hep-th/9301042  }

\bibitem{DineSeib}{M.Dine, N.Seiberg,
" Is the superstring weakly coupled?", Phys.Lett.B162(1985)299-302;
M.Dine, R.Rohm, N.Seiberg, E. Witten, Phys.Lett.B156(1985)55;
M.Dine, N.Seiberg, Phys.Rev. Lett.55(1985)366}
\bibitem{Townsend}{P.Townsend, Brane Surgery Nucl. Phys. B., 
Proc. Suppl. 58 (1997) 163-175.[HEP-TH 9609217]}
\bibitem{branes ending on branes}{A. Strominger Open P-Branes, 
Phys.Lett.B383:44-47,1996 ;hep-th/9512059 }
\bibitem{Szg&Howe}{P.S. Howe, E.Sezgin, Superbranes,
Phys.Lett. B390 (1997) 133-142;hep-th/9607227;  D=11, p=5, Phys.Lett. B394
(1997) 62-66;hep-th/9611008;
 P.S. Howe, E. Sezgin, P.C. West, Covariant Field Equations of 
the M Theory Five-Brane, Phys.Lett. B399 (1997) 49-59;hep-th/9702008 }

\bibitem{Sorokin}{ D.Sorokin,Superbranes and Superembeddings, Phys.Rept.
329 (2000) 1-101 (hep-th/9906142)}
\bibitem{11tor}{E. Cremmer and S. Ferrara, Phys. Lett. {\bf 91B} (1980)
61;\\
L. Brink and P. S. Howe,  Phys. Lett. {\bf 91B} (1980) 384.}
\bibitem{DWPP}{B. de Wit, K. Peeters, and J. Plefka, Superspace 
geometry for supermembrane backgrounds,'' Nucl.Phys. B532 (1998) 99; 
}
\bibitem{shibusa}{Y. Shibusa,
 ``11-dimensional curved backgrounds for supermembrane in superspace,''}
\bibitem{Sezgin}{ E. Sezgin, P. Sundell,'' Aspects of the M5-Brane'',
  hep-th/9902171     }
\bibitem{Fab}{M.Fabinger, P.Ho\v rava,''Casimir effect between
world-branes
in heterotic M-theory, hep-th/0002073}
\bibitem{PST}{{P. Pasti, D. Sorokin and M. Tonin, Physics
Lett.{B398}{97}{41}, 
(hep-th/9701037); \\
I. Bandos, K. Lechner, A. Nurmagambetov, P. Pasti, D. Sorokin and 
M. Tonin, Phys. Rev. Lett.{78}{97}{4332}, (hep-th/9701149).}}
\bibitem{Sch}{
        M. Aganagic, J. Park, C. Popescu, J. Schwarz,
``World-Volume Action of the M Theory Five-Brane'',
        Nucl.Phys. B496 (1997) 191-214;hep-th/9701166 }
\bibitem{WB}{
J.Wess and J. Bagger,Supersymmetry and Supergravity,Second ed.,Princeton
University Press 1992.
}
\bibitem{cremmer}{ E. Cremmer, S. Ferrara, L. Girardello, A. Van Proeyen,
 `` Yang-Milles theories with local supersymmetry: Lagrangian,
transformation
laws and superhiggs effect,''
 Nucl.Phys.B212:413,1983 
}
\bibitem{Der}{J.P.Derendinger, R.Sauser,
 ``A five-brane Modulus in the Effective N=1 Supergravity of M-theory'',
hep-th/0009054}
\bibitem{Brax}{Ph.Brax, J.Mourad,
'' Open Supermembranes Coupled to M-Theory Five-Branes'',
       Phys.Lett. B416 (1998) 295-302; hep-th/9707246 }
\bibitem{perelomov}
{Perelomov.A,  Integrable systems of classical mechanics and Lie algebras.
Birkhuser Verlag, 1990. 
}
\bibitem{smallinst}{
B.Ovrut, T.Pantev, J.Park: Small Instanton Transitions in Heterotic
M-Theory,
JHEP 0005 (2000) 045; hep-th/0001133  }

\bibitem{munoz}{
C. Munoz, ``Effective Supergravity from Heterotic M-Theory and its
Phenomenological Implications,'' 
hep-th/9906152; D.G. Cerdeno and C. Munoz, ``Phenomenology of Non-Standard 
Embedding and Five-branes in M-Theory,'' hep-ph/9904444} 

\bibitem{BST}{ E. Bergshoeff, E. Sezgin, and P.K. Townsend,
``Supermembranes and 11-dimensional supergravity,''
Phys. Lett. {\bf 189B}(1987)75;
``Properties of the Eleven-Dimensional Supermembrane
Theory,'' Ann. Phys. {\bf 185}(1988)330.}

\bibitem{GSW}{M. Green, J.H.  Schwarz, and E. Witten, {\it Superstring 
Theory}, Cambridge University Press, Cambridge (1987)}

\bibitem{Li}{T.Li, J.L.Lopez and D.V.Nanopoulos,
 Phys.Rev.D56,2602,(1997)}
\bibitem{Dud}{E.Dudas,C.Grojean, Nucl.Phys.B507,553 (1997)}
\bibitem{Nil}{H.P.Nilles, M.Olechowski, M.Yamaguchi, Phys.Lett.B415,24
(1997);
Nucl.Phys.B530,43 (1998) }
\bibitem{Stieb}{ S. Stieberger,'' (0,2) Heterotic Gauge Couplings and
 their M-Theory Origin'',
        Nucl.Phys. B541 (1999) 109-144}
\bibitem{Kim}{K. Choi, H. B. Kim, H. Kim,
`` Moduli Stabilization in Heterotic $M$-theory'',
 Mod.Phys.Lett. A14 (1999) 125-134, 
hep-th/9808122 }
\bibitem{Benakli}{K. Benakli,
`` Phenomenology of Low Quantum Gravity Scale Models,''
        Phys.Rev. D60 (1999) 104002}
\bibitem{modcosmo}{T. Banks, M. Berkooz, G. Moore, S.H. Shenker, and 
P. Steinhardt, ``Modular Cosmology,'' hep-th/9503114; 
Phys.Rev. D52 (1995) 3548-3562}
\bibitem{Vafa1}{M. Atiyah, J. Maldacena, C. Vafa, 
`` An M-theory Flop as a Large N Duality,''hep-th/0011256 }
\bibitem{Vafa2}{ C. Vafa,''Superstrings and Topological Strings at 
Large N'',
hep-th/0008142 }
\bibitem{Vafa3}{ H. Ooguri, C. Vafa, 
`` Knot Invariants and Topological Strings'',Nucl.Phys. B577 (2000) 419-438}
\bibitem{Kachru1}{S. Kachru, S. Katz, A. Lawrence, J. McGreevy,
``Mirror symmetry for open strings'', Phys.Rev. D62 (2000) 126005,
 hep-th/0006047}
\bibitem{Kachru2}{S.Kachru, E.Silverstein,
``Chirality Changing Phase Transitions in 4d String Vacua'',
Nucl.Phys. B504 (1997) 272-284, hep-th/9704185}
\bibitem{CandGreen}{P.Candelas, X. C. de la Ossa. P.Green,
L.Parkes,''A pair of Calabi-Yau Manifolds as an exactly
soluble superconformal theory'', Essays on Mirror Manifolds,
ed. S.T.Yau} 
\bibitem{AspMorr}{P.S. Aspinwall and D.R. Morrison, 
``Topological field theory and rational curves,'' 
Commun. Math. Phys. {\bf 151}(1993)245}
\bibitem{Marino}{ M. Marino, G. Moore,
``Counting higher genus curves in a Calabi-Yau manifold'',
 Nucl.Phys. B543 (1999) 592-614,
hep-th/9808131} 
\bibitem{faber}{C. Faber and R. Pandharipande, ``Hodge 
Integrals and Gromov-Witten Theory,'' math.AG/9810173}
\bibitem{Vafa4}{ R. Gopakumar, C. Vafa,
`` M-Theory and Topological Strings--I'',hep-th/9809187;
 ``M-Theory and Topological Strings--II'', hep-th/9812127} 
\bibitem{Morrison}{D. R. Morrison, M. R. Plesser,
`` Summing the Instantons: Quantum Cohomology and Mirror Symmetry 
in Toric Varieties'', Nucl. Phys. B440 (1995) 279-354,
hep-th/9412236 }
\bibitem{DonagiOW}{R. Donagi, B.A. Ovrut, and D. Waldram, 
``Moduli spaces of fivebranes on elliptic Calabi-Yau threefolds,'' 
hep-th/9904054}
\end{thebibliography}
\end{document}